\documentclass[a4paper,fleqn,usenatbib]{mnras}

\usepackage{newtxtext,newtxmath}

\usepackage[T1]{fontenc}
\usepackage{ae,aecompl}

\usepackage{graphicx}	
\usepackage{epstopdf}
\usepackage{amsmath}	

\def\VEV#1{\left\langle #1\right\rangle} 
\newcommand{\bq}{\begin{equation}}
\newcommand{\eq}{\end{equation}}
\newcommand{\bqa}{\begin{eqnarray}}
\newcommand{\eqa}{\end{eqnarray}}

\usepackage{etoolbox}

\title[Bursts during the Cosmic Dawn]{Bursty star formation during the Cosmic Dawn driven by delayed stellar feedback}

\author[Furlanetto \& Mirocha]{
Steven R.~Furlanetto$^{1}$\thanks{E-mail: sfurlane@astro.ucla.edu} \& Jordan Mirocha$^2$ \\
$^{1}$Department of Physics \& Astronomy, University of California, Los Angeles, Los Angeles, CA 90095, USA\\
$^{2}$Department of Physics and McGill Space Institute, McGill University, 3600 University Street, Montreal, QC H3A 2T8, Canada}
\date{Accepted XXX. Received YYY; in original form ZZZ}

\pubyear{2015}

\begin{document}
\label{firstpage}
\pagerange{\pageref{firstpage}--\pageref{lastpage}}
\maketitle

\begin{abstract}
In recent years, several analytic models have demonstrated that simple assumptions about halo growth and feedback-regulated star formation can match the (limited) existing observational data on galaxies at $z \ga 6$. By extending such models, we demonstrate that imposing a time delay on stellar feedback (as inevitably occurs in the case of supernova explosions) induces burstiness in small galaxies. Although supernova progenitors have short lifetimes ($\sim 5$--30~Myr), the delay exceeds the dynamical time of galaxies at such high redshifts. As a result, star formation proceeds unimpeded by feedback for several cycles and ``overshoots" the expectations of feedback-regulated star formation models. We show that such overshoot is expected even in atomic cooling halos, with masses up to $\sim 10^{10.5} \ M_\odot$ at $z \gtrsim 6$. However, these burst cycles damp out quickly in massive galaxies, because large haloes are more resistant to feedback so retain a continuous gas supply. Bursts in small galaxies -- largely beyond the reach of existing observations -- induce a scatter in the luminosity of these haloes (of $\sim 1$~mag) and increase the time-averaged star formation efficiency by up to an order of magnitude. This kind of burstiness can have substantial effects on the earliest phases of star formation and reionization.
\end{abstract}




\section{Introduction} \label{intro}

Galaxy evolution is one of the most fundamental 
areas of astrophysics and is a prime focus on both observers and theorists. However, the tangle of processes contributing to it -- across an enormous range of scales, from cosmological structure formation to small-scale star formation -- also makes it one of the most challenging. Galaxy evolution during the ``Cosmic Dawn" at $z \ga 6$ is a particular puzzle, both because there are relatively few observations of galaxies during this era and because theorists expect a host of new physics (such as large-scale radiative feedback and exotic stellar populations) to be important.

Nevertheless, over the past several years, models have shown that existing observations of (relatively bright) $z > 6$ galaxies can be explained through a simplified framework \citep{trenti10, tacchella13, tacchella18, mason15, sun16, furl17-gal} that extrapolates the basic physical processes most relevant for star-forming galaxies at later times \citep{bouche10, dave12, dekel13, lilly13}. These models typically assume that star formation is fueled by cosmological gas accretion onto the host dark matter halo, with the star formation rate determined by the efficiency of stellar feedback. Simple estimates of feedback provide reasonable fits to the observed luminosity functions of galaxies in this early era. 

However, the agreement is certainly not perfect. For example, the most basic models of feedback rely on balancing the energy or momentum output from stellar radiation and/or supernovae with the binding energy of the host halo. Because haloes form at higher densities in the early Universe, this prescription implies more efficient star formation at earlier times. Fits to the observed luminosity functions (which only sample fairly bright galaxies during this era) mildly prefer a redshift-independent star formation efficiency (e.g., \citealt{mirocha17}). Redshift-dependent models of the star formation process generally require non-trivial changes in the dust properties \citep{Yung2019a,Vogelsberger2019,Qiu2019} and/or duty cycle of star formation \citep{mirocha20} in order preserve agreement with observations. This issue (amongst others) motivates continued work into a more detailed understanding of these early sources. 

Recently, \citet{furl20-disc} explored more complex models of galaxy formation at $z>6$, demonstrating that even with more sophisticated treatments of star formation, the overall strength of galactic feedback still controls the star formation efficiency. However, these models made the key assumption of ``quasi-equilibrium" feedback regulation:  star formation and the accompanying stellar feedback were assumed to occur at precisely the same time. Of course, this is not an accurate assumption in detail, as there is a delay of $\sim 5$--30~Myr between the birth of massive stars and their supernovae.  \citet{faucher18} pointed out that this timescale becomes significant in the early Universe, because the dynamical time (which ultimately controls the rate at which gas can collapse to high densities and form stars) $t_{\rm dyn} \propto (1+z)^{-3/2}$.  Even at later times, when the dynamical time is quite long, \citet{orr19} showed that delayed feedback can induce oscillations in the star formation history.

Indeed, at $z \ga 6$, the star formation timescale is comparable to the feedback delay, which suggests that this simple delay cannot be ignored in galaxy models. Intuitively, one then expects that star formation will occur through repeated bursts, with episodes running away until feedback kicks in and temporarily shuts off star formation (or at least decreases its rate).  Detailed numerical simulations, which typically incorporate stellar feedback in this manner, show strong bursts in both early galaxies (e.g., \citealt{kimm15}) and dwarfs (e.g., \citealt{weisz12, broussard19, wheeler19, najmeh19, chavesmontero21, iyer20, kravtsov21}), though other forms of feedback can moderate these effects (e.g., \citealt{smith21}). 

However, such detailed models also incorporate additional physics, including other mechanisms that may trigger bursty star formation (including the stochastic nature of halo accretion and the small size of some early galaxies) or suppress it. In this paper, we take a different approach. We incorporate delayed feedback into our analytic model of galaxy formation (building upon \citealt{furl17-gal, mirocha17, furl20-disc}) and examine its implications for galaxies in the early Universe. We find that this simple change can have profound effects on the overall course of star formation in this era.

This paper is organized as follows. In section \ref{simple-bursts}, we introduce a toy model for bursts that motivates many of the key results. Then we introduce the model for delay-driven bursts in section \ref{burst-model}, and we examine its implications for individual galaxies and galaxy populations in sections \ref{ind-gal} and \ref{burst-pops}, respectively. We consider some consequences of bursts for future observations and reionization in section \ref{discussion}, and we conclude in section \ref{conc}.

The numerical calculations here assume $\Omega_m = 0.308$, $\Omega_\Lambda = 0.692$, $\Omega_b = 0.0484$, $h=0.678$, $\sigma_8=0.815$, and $n_s=0.968$, consistent with the recent results of \citet{planck18}. 

\section{A Simple Model for Bursts}
\label{simple-bursts}

In this section, we consider a very simple toy model for repeated bursts of star formation, which will illustrate some of the key takeaways from the full model. 

\subsection{The quasi-equilibrium model} \label{eq-model}

To begin, let us consider a model \emph{without} bursts, in order to establish a baseline against which we can isolate the effects of delayed feedback and bursts. This model is a simplified version of the ``bathtub" model \citep{bouche10, dave12, dekel14}. Let us imagine that a galaxy accretes material, forming stars with a fixed efficiency per free fall time $\epsilon_{\rm ff}$. We let $t_{\rm ff}$ be the system's free fall time, which we assume to be constant for simplicity. The newly-formed massive stars exert feedback on the surrounding gas, ejecting it with a (constant) efficiency $\eta = \dot{m}_w/\dot{m}_\star$, where $\dot{m}_w$ is the rate at which material is ejected from the system.  Finally, we assume that the system accretes gas from its surroundings at a constant rate $\dot{m}_{c,g}$.\footnote{In reality, all of these parameters will evolve as the galaxy grows, of course, but they will typically not change much over one burst cycle. We therefore neglect such variations for the toy  model.} Then, the stellar mass and gas mass of the system evolve according to:
\bqa
\dot{m}_\star & = & {\epsilon_{\rm ff} \over t_{\rm ff}} m_g \\
\dot{m}_g & = & \dot{m}_{c,g} - \dot{m}_\star - \eta \dot{m}_\star.
\eqa
We non-dimensionalize these equations by transforming to a time variable $u = t/t_{\rm ff}$ and a mass variable $\tilde{m} = m/m_{\rm acc}$, where $m_{\rm acc} = \dot{m}_{c,g} t_{\rm ff}$. Then we obtain
\bqa
\tilde{m}'_\star & = & \epsilon_{\rm ff} \tilde{m}_g \\ 
\tilde{m}'_g & = & 1 - \epsilon_{\rm ff} \tilde{m}_g - \eta \epsilon_{\rm ff} \tilde{m}_g, \label{eq:gas-bathtub}
\eqa
where primes denote derivatives with respect to $u$.

After an initial transient phase, this simple model reaches a quasi-equilibrium state in which the star formation rate is set by the condition that feedback expels all of the accreting gas that is not transformed into stars. This has the solutions
\bqa
\tilde{m}_g & = & { 1 \over \epsilon_{\rm ff} (1 + \eta)} \left[ 1 - e^{- \epsilon_{\rm ff} (1 + \eta) u} \right] \\
X_\star^{\rm eq} & \rightarrow & {1 \over 1 + \eta} \qquad \qquad \qquad u \rightarrow \infty,
\eqa
where the latter applies after the initial transient phase ($\epsilon_{\rm ff} (1 + \eta) u \gg 1$).

Importantly (and perhaps counter-intuitively), in this regime the net star formation efficiency is independent of the parameter $\epsilon_{\rm ff}$, depending only on the feedback strength $\eta$: the gas reservoir adjusts so as to produce stars at the rate required for feedback to balance accretion. This behavior also occurs in models with more sophisticated star formation and feedback prescriptions (e.g., \citealt{furl20-disc}), but we will see that bursts behave very differently.

\begin{figure*}
	\includegraphics[width=\columnwidth]{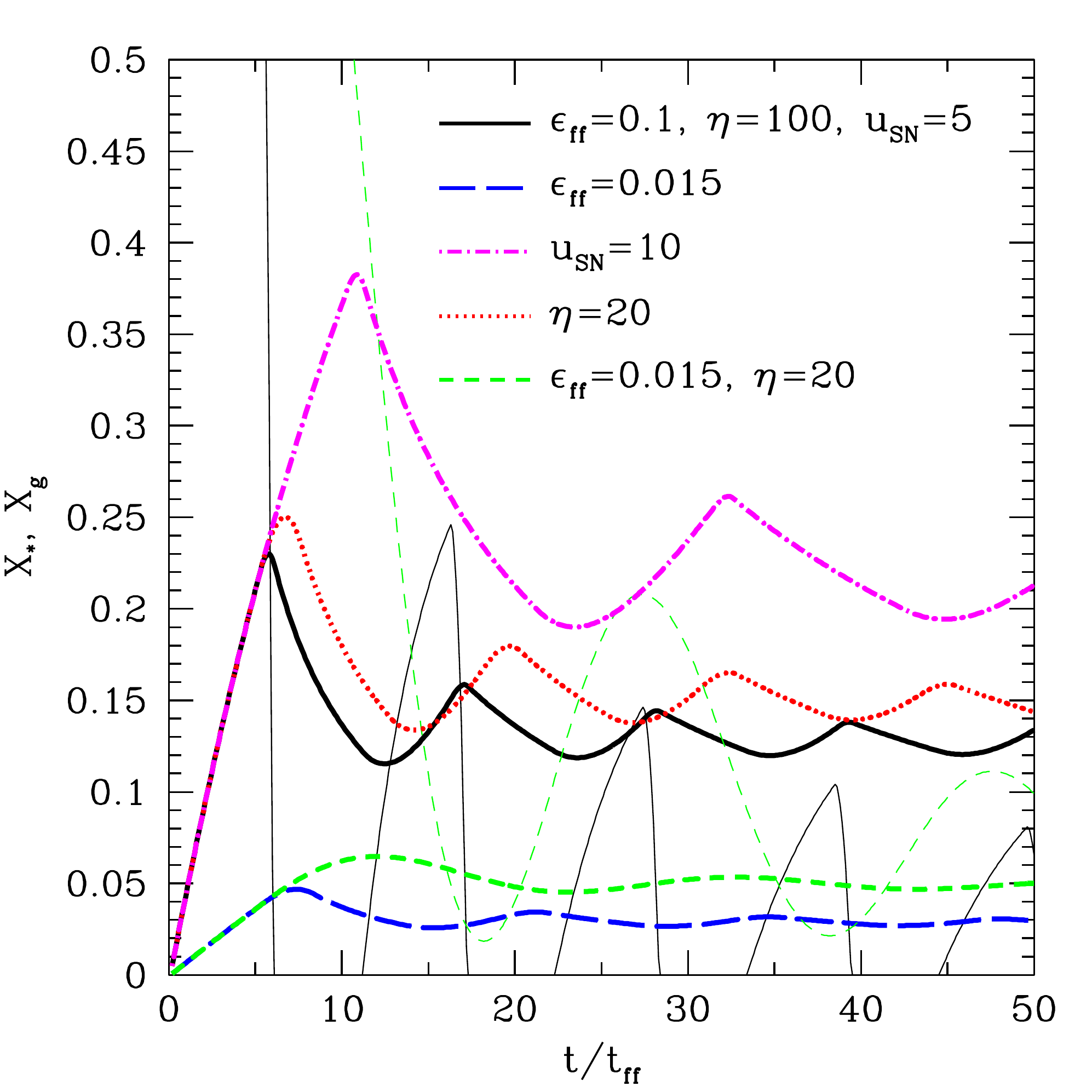} \includegraphics[width=\columnwidth]{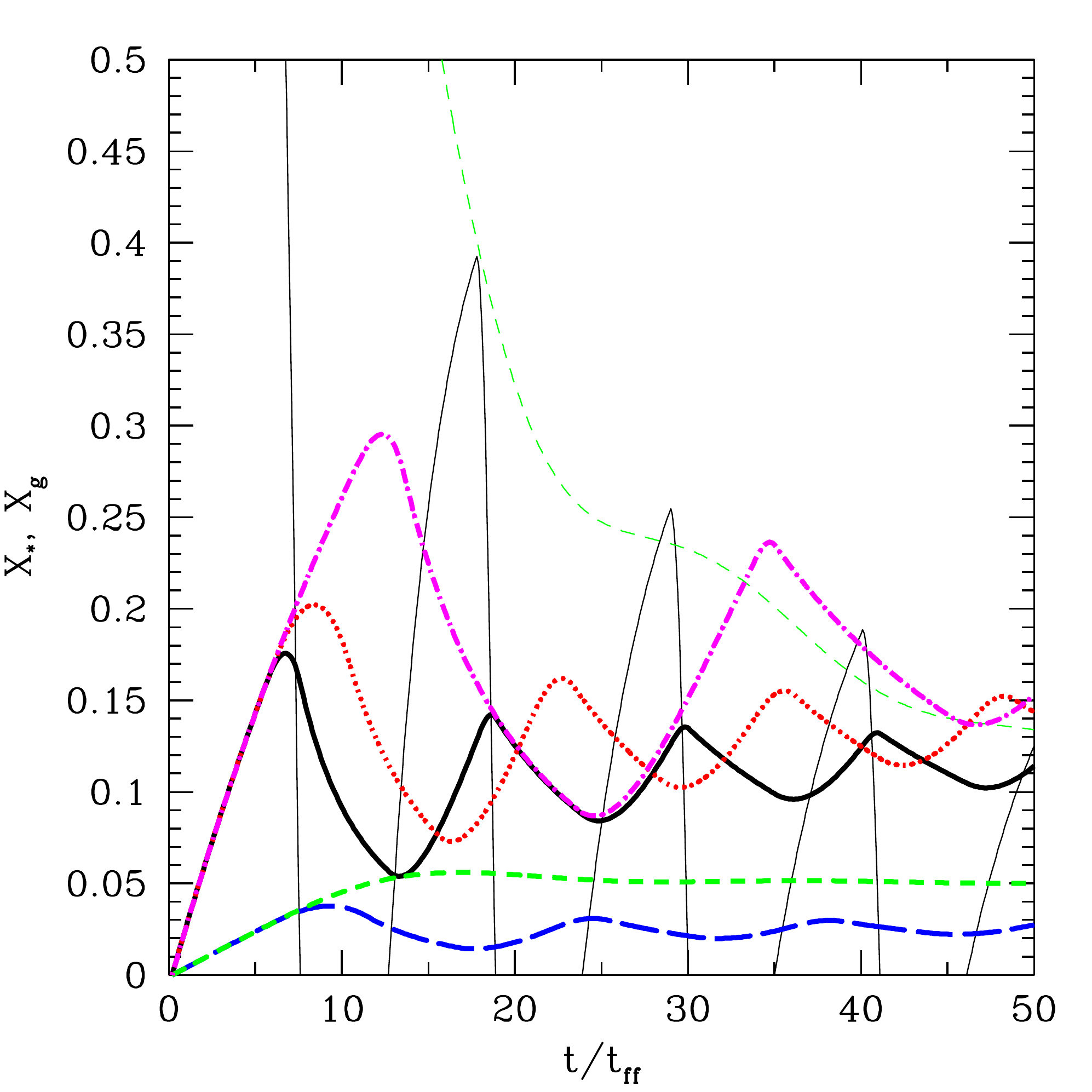} 
    \caption{Toy models of feedback-driven bursts. The left panel shows models in which the accretion rate is constant with time, while in the right panel it increases with halo mass. The thick curves show the stellar mass fraction $X_\star = m_\star/m_a$, while the thin curves show the corresponding gas fractions for two models. The solid curve is our fiducial model, with $u_{\rm SN} = 5$, $\epsilon_{\rm ff}=0.1$, and $\eta=100$. The other curves take the same parameters except $\eta=20$ (dotted curve), $u_{\rm SN}=10$ (dot-dashed curve), $\epsilon_{\rm ff}=0.015$ (long-dashed curve), and both $\eta=20$ and $\epsilon_{\rm ff}=0.015$ (short-dashed curve). }
    \label{fig:toy-models}
\end{figure*}

\subsection{A simple burst}

Next we introduce a simple change to the bathtub model: we assume that feedback is delayed, being injected only after a time $t_{\rm SN}$ has passed. Again for simplicity, we assume that feedback from any episode of star formation is injected instantaneously after this fixed delay time. Equation~(\ref{eq:gas-bathtub}) then becomes
\bq
\tilde{m}'_g = 1 - \epsilon_{\rm ff} \tilde{m}_g - \eta \epsilon_{\rm ff} \tilde{m}_g^d, \label{eq:gas-delay}
\eq
where primes denote derivatives with respect to $u$ and $\tilde{m}_g^d$ is the gas mass at the delayed ``time" coordinate $u - u_{\rm SN}$, where $u_{\rm SN} \equiv t_{\rm SN}/t_{\rm ff}$.

The solution to this system is complicated by the delay term, but an analytic solution can still be obtained in a piecewise fashion. For $u < u_{\rm SN}$, the feedback term vanishes, because no supernovae have occurred yet. In that period, the solution is straightforward,
\bqa
\tilde{m}_g & = & \epsilon_{\rm ff}^{-1} [ 1 - e^{-\epsilon_{\rm ff} u} ] \qquad (u \le u_{\rm SN}) \\
\tilde{m}_\star & = & u - \tilde{m}_g. 
\eqa
Put simply, the gas is transformed into stars at a constant efficiency in the absence of feedback, with the overall star formation rate increasing as the gas reservoir grows through accretion. We can then insert this solution for $\tilde{m}_g$ into the feedback term in equation~(\ref{eq:gas-delay}), because that feedback depends upon stars formed during the ``unperturbed" early star formation. The resulting equation is integrable, with the solution in terms of $w = u - u_{\rm SN}$
\bqa
\tilde{m}_g & = & \tilde{m}_g(u_{\rm SN}) e^{-\epsilon_{\rm ff} w} + { (1 - \eta) \over \epsilon_{\rm ff}} \left[ 1 - e^{-\epsilon_{\rm ff} w} \right] \nonumber \\
& & + \eta w e^{-\epsilon_{\rm ff} w} \\
\tilde{m}_\star & = & \tilde{m}_\star(u_{\rm SN}) + \tilde{m}_g(u_{\rm SN} ( 1 - e^{-\epsilon_{\rm ff} w}) \nonumber \\
& & + { (1 - \eta) \over \epsilon_{\rm ff} } \left[ \epsilon_{\rm ff} w + ( e^{-\epsilon_{\rm ff} w} - 1) \right] \nonumber \\ 
& & + {\eta \over \epsilon_{\rm ff}} \left[ 1 - e^{-\epsilon_{\rm ff} w} ( 1 + \epsilon_{\rm ff} w) \right].
\eqa

One can continue with the piecewise solution, but this first phase provides the necessary intuition. It is useful to present the results in terms of the fraction of the accreted gas (which has a total mass $m_a$) that is transformed into stars, $X_\star \equiv m_\star/m_a = \tilde{m}_\star/u$. The left panel of Figure~\ref{fig:toy-models} shows some of these solutions over several cycles, with several different parameter combinations. The thick curves show $X_\star$, while the dotted curves show (for two of the cases) the corresponding fraction in the gaseous phase, $X_g = m_g/m_a$. The solid curve uses our fiducial parameters ($u_{\rm SN} = 5$, $\epsilon_{\rm ff}=0.1$, and $\eta=100$), while the others vary these parameters to illustrate their importance.

In all cases, the stellar fraction increases rapidly until $t_{\rm SN}$, unimpeded by feedback. After that time, supernova feedback slows or stops star formation, and $X_\star$ decreases. In the fiducial model, the thin solid curve shows that the gas reservoir is very quickly evacuated after $t_{\rm SN}$, because the combination of efficient star formation and feedback causes the system to ``overshoot" the equilibrium expectation. Star formation in this regime does not resume until $ \approx 2t_{\rm SN}$, when feedback from that initial burst shuts off.  The cycle then repeats. While each burst episode is roughly the same as the previously ones, the fluctuations in $X_\star$ damp out because the halo's mass grows steadily. (Note that our assumption of a constant $t_{\rm ff}$ will break down in the cosmological case, which we explore in the next section.)

Most of the other models shown here behave qualitatively similarly -- we see, for example, that slowing the feedback (dot-dashed curve) or decreasing its strength (dotted curve) increase the duration of each burst and so increase the net star formation efficiency $X_\star$, although the effect of $\eta$ is quite modest. However, the model with weak star formation and feedback ($\epsilon_{\rm ff}=0.015$, and $\eta=20$; short-dashed curve) behaves quite differently. Here feedback from the initial burst is not strong enough to eject the gas reservoir completely, so the system settles into fairly steady growth.

Pronounced bursts will occur in the limit that the outflow rate per time step exceeds the accretion rate, 
$\epsilon_{\rm ff} \eta > 1$. In this case, the initial (pre-feedback) burst of star formation can evacuate the gas reservoir. We can solve for the time $u_{\rm ev}$ when $\tilde{m}_g =0$, at which point the initial burst will have burnt itself out. In the limit in which the stellar mass is much smaller than the accreted mass, $\epsilon_{\rm ff} u \ll 1$, this time is
\bq
u_{\rm ev} = u_{\rm SN} + {1 \over \epsilon_{\rm ff} \eta} \left[ 1 + \sqrt{2 \epsilon_{\rm ff} \eta u_{\rm SN}} \right].
\eq
In the additional limit of strong feedback $\epsilon_{\rm ff} \eta \gg 1$, the gas is evacuated shortly after the supernovae become active, and $u_{\rm ev} \approx u_{\rm SN}$, but for more moderate cases star formation can extend some ways past the onset of feedback (as in the dotted curve). This helps to compensate for a low star formation efficiency and makes the overall star formation efficiency less sensitive to that parameter during a bursty phase.

We can also estimate the stellar mass fraction at the end of the burst, when $u=u_{\rm ev}$:
\bq
X_\star(u_{\rm ev}) \approx {1 \over 2} \left[ {1 \over \eta} + \sqrt{2 \epsilon_{\rm ff} u_{\rm SN} \over \eta} + \epsilon_{\rm ff} u_{\rm SN} \right].
\eq
However, the halo continues to grow when stars are unable to form. The net star formation efficiency after a complete cycle (just before stars can form again) is
\bq
X_\star^{\rm net} \approx {\epsilon_{\rm ff} \over 2} {u_{\rm ev}^2 \over u_{\rm ev} + u_{\rm SN}}.
\label{eq:xnet-anal}
\eq
If the stellar feedback from the initial burst is strong, or $\epsilon_{\rm ff} \eta u_{\rm SN} \gg 1$, then $X_\star^{\rm net} \rightarrow \epsilon_{\rm ff} u_{\rm SN}/4$. But, again, weaker feedback allows star formation to persist longer during the burst, so that the dependence on $\epsilon_{\rm ff}$ moderates.\footnote{Note as well that this expression modestly overestimates $X_\star$ in the toy model because it ignores the gas removed from the reservoir by star formation and by feedback during the interval from $u_{\rm SN}$ to $u_{\rm ev}$. }

Now we can see that, in this strong burst limit, the stellar mass fraction is insensitive to the feedback parameter $\eta$ but does depend on the small-scale star formation efficiency $\epsilon_{\rm ff}$. This stands in stark contrast to the equilibrium models described in section \ref{eq-model}, in which the solution is independent of $\epsilon_{\rm ff}$ but strongly dependent on $\eta$. 

\subsection{Cosmological halo growth}

In a cosmological context, the previous model ignores the fact that the system's key parameters, and especially the free-fall time, evolve as the galaxy grows. This adds an important wrinkle to the solution. Most importantly, the average accretion rate for haloes is approximately $\dot{m}_{c,g} \propto m_h$ \citep{dekel13}. Our non-dimensional system has scaled out the mass accreted per free-fall time, but we can see the qualitative effects of the increasing accretion rate by setting $\dot{m}_{c,g} \propto m_a$. In this case, later bursts have more fuel so form more stars. Thus we expect the oscillations in $X_\star$ to damp out more slowly with time (and the oscillations in the absolute star formation rate to grow).

The right panel of Figure~\ref{fig:toy-models} shows some of these solutions, using the same parameter choices as in the left panel. As expected, the fluctuations take longer to ``damp out." We also see that the peak $X_\star$ values are systematically smaller, but this is just because the halo is growing faster overall. However, for the short-dashed curve -- which has weak feedback and inefficient star formation -- the gas is never evacuated, and because of the increasing accretion rate it reaches the steady state, in which $X_\star \approx (1 + \eta)^{-1}$ \emph{faster}.

Over cosmological timescales, we will also have non-trivial redshift dependence to other parameters: the accretion rate $\propto (1+z)^{5/2}$, while the free-fall time $t_{\rm ff} \propto (1+z)^{-3/2}$. At early times, there will therefore be both higher accretion rates and more star formation cycles per supernova feedback interval. For both of these reasons, we therefore expect more ``overshoot" due to burstiness early on. 

\subsection{A star formation threshold} \label{toy-threshold}

We have ignored another potential aspect of star formation inside these galaxies: the accreted gas may not be immediately available for star formation. Consider the simple picture of a galaxy disc undergoing cosmological accretion. For such a system, stability is determined by the balance between stellar feedback (which helps support the disc) and gravitational fragmentation (which drives star formation). Over long timescales, galaxies approach a quasi-equilibrium in which these two processes (as well as other sources of disc support) balance \citep{thompson05, munoz12, faucher13, krumholz18}, so that the gas disc remains on the cusp of gravitational fragmentation (see section \ref{disc-burst}).

When $t_{\rm ff} \ll t_{\rm SN}$, however, it may be difficult to reach that marginally-stable condition. In that case, gas could build up until the surface density is high enough to drive fragmentation, at which point star formation would begin and continue unimpeded for a time $t_{\rm SN}$. Then the feedback begins, overshooting the stability condition and evacuating the gas reservoir entirely. The next star formation cycle will be delayed compared to the toy model, because the gas reservoir must build itself up until it reaches the point of instability again.

We can explore the qualitative impact of such a scenario simply by requiring that star formation does not occur until the gas mass exceeds a threshold $m_{\rm sf}$. 
In our simple model, we can incorporate the threshold simply by shifting the origin of the time coordinate $u$ to be be at the point where the accretion has just exceeded the threshold, which we will call $\tilde{m}_g^{i}$. Then we must simply adjust our solutions to include this gas reservoir in the initial conditions. The solutions differ significantly in the limit $\tilde{m}_g^{i} \gg u_{\rm SN}$ (or in other words if the threshold mass is much larger than the amount of gas accreted before feedback begins), because in that case the star formation rate remains fairly steady throughout the burst phase (with $\tilde{m}'_\star \approx \epsilon_{\rm ff} \tilde{m}_g^{i}$) rather than growing significantly over $u_{\rm SN}$ as the halo accretes material. As a result, in this limit the stellar mass fractions behave somewhat differently, with, for example,
\bq
X_\star^{\rm pk} \approx \sqrt{\epsilon_{\rm ff} \tilde{m}_g^i} \left( \sqrt{1 \over 2 \eta} + \sqrt{\epsilon_{\rm ff} \tilde{m}_g^i} \right).
\eq

In short, imposing a threshold increases the time-averaged star formation efficiency, because throughout the burst period the reservoir of star-forming gas is large. The enhancement is of order unity, however, so we expect the details of the star formation model to be relatively unimportant to the burst solutions. We will see that this is indeed the case by examining two different burst models in the next section.

\section{A burst model for early galaxies} \label{burst-model}

We now describe a model that permits bursty star formation in early halos, driven by the significant time delay between star formation and supernova feedback. This framework will by no means provide a complete model of galaxy formation, but it will serve to highlight the potential contributions of a burst mode to early star formation. We will see many of the same features as the toy model. 

\subsection{Halo assembly}

We assume that haloes grow in a manner that maintains constant number density over time \citep{furl17-gal}, which has been shown to match the average halo growth in numerical simulations fairly well \citep{mirocha21}. We use the \citet{trac15} mass functions to determine the number density at each redshift, as it has been tested at high redshifts. For intuition, this is not too far from the simple relation inspired by simulations \citep{dekel13},
\bq
\dot{m}_h \approx A m_h (1+z)^{5/2}
\label{eq:macc}
\eq
with a corresponding gas accretion rate $\dot{m}_{c,g} = (\Omega_b/\Omega_m) \dot{m}_h$. 

\subsection{The evolution equations}

The star formation model has three key components: (1) gas accretes onto a reservoir; (2) stars form at a specified rate from the gas reservoir; and (3) after a delay, the stars inject momentum into the gas, ejecting some gas from the system. (In section \ref{disc-burst} we will supplement these with a model for the stability of the reservoir, assuming it is a disc.) We use this simple system of equations,
\bqa
\dot{m}_g & = & \dot{m}_{c,g} - \dot{m}_\star - \dot{m}_w, \\
\dot{m}_\star & = & {\epsilon_{\rm ff} \over t_{\rm ff}} f_{\rm sf} m_g, \label{eq:sf-evol} \\
\dot{m}_w & = & \eta \dot{m}_\star^d. 
\eqa
The first equation describes the gas reservoir, with gas added through cosmological accretion and removed both through star formation and feedback. The second describes star formation: we assume that a fraction $f_{\rm sf}$ of the gas is eligible to form stars, and a fraction $\epsilon_{\rm ff}$ of this material forms stars per free-fall time in the galaxy. Gas is ejected with a mass-loading factor $\eta$, based on an effective star formation rate $\dot{m}_\star^d$ that accounts for the delay. 

We will now discuss the latter two equations in more detail.

\subsection{Star formation}

Although equation~(\ref{eq:sf-evol}) for the star formation rate appears simple, it hides some important complexity. The star formation efficiency $\epsilon_{\rm ff}$ depends on the details of star formation on small scales. In local molecular clouds, it takes the roughly constant value $\epsilon_{\rm ff} \approx 0.015$ (e.g. \citealt{krumholz12,salim15, leroy17}, though see \citealt{murray11, lee16}), but there is no particular reason to suppose that value would apply to the first galaxies. We therefore treat this as a free parameter. (As we have already discussed in section \ref{eq-model}, when galaxies reach their quasi-equilibrium marginally stable state, $\epsilon_{\rm ff}$ has only a small effect on a galaxy's star formation rate.) 

The galaxy's free-fall time will depend on the density reached by the gas, but it will be related to the orbital timescale of the gas elements, $t_{\rm ff} \propto (G \bar{\rho})^{-1/2} \sim t_{\rm orb}$.  We set the fiducial proportionality constant by assuming that the gas settles into a disc that is marginally unstable to gravitational collapse, in which case \citep{faucher18},\footnote{Below we will consider a model in which we assume that star formation occurs in such a disc, but for now we just use that geometry to set the fiducial value of the free-fall time.}
\bq
t_{\rm ff} \sim 0.2 t_{\rm orb},
\eq
where $t_{\rm orb} = 2 \pi r/v = 1/\Omega$, $r$ is the orbital radius, $v$ is the orbital velocity, and $\Omega$ is the orbital frequency. To estimate all of these, we use the scaling relations from \citet{faucher18}, which assume discs form inside NFW haloes, and we evaluate characteristic quantities at the half-mass radii. The free-fall time only appears in our equations in combination with $\epsilon_{\rm ff}$, so we absorb uncertainties in the timescale into the efficiency parameter. We note also that the orbital timescale is independent of halo mass at a fixed cosmic time, because $t_{\rm orb} \propto r_{\rm vir}/v_c \propto (1+z)^{-3/2}$.

The remaining factor is the fraction of gas available to star formation. We do not attempt to model this in any detail here. Instead, for now we set $f_{\rm sf}=1$. We will explore the importance of this parameter in section \ref{disc-burst}. 

\subsection{Outflows}

The last crucial component of our model is feedback, which requires two physical inputs. First, we must determine the timing of the feedback injection. Given an initial mass function for stars, a model for stellar lifetimes, and the mass range of stars that end in supernova explosions, one can specify this injection rate precisely. However, other feedback processes (such as radiation) may also be relevant. For simplicity, we assume that feedback generated by an episode of star formation is injected uniformly over an interval $t_{\rm d,min} < t < t_{\rm d,max}$. As a fiducial choice, we assume this range corresponds to $5$--30~Myr. (Note that \citealt{orr19} find that, for a typical IMF, the rate of momentum injected by feedback decreases with time as $\sim t^{-1/2}$ over that period.)

Second, we must determine the efficiency with which material is expelled from the gas reservoir. This requires assumptions about the energy available to drive the outflow (and in particular how much of it is lost through radiative cooling) as well as a model for how the energy and momentum is distributed throughout the gas reservoir. 
In our fiducial model, we obtain $\eta$ by assuming that feedback accelerates gas to the halo's escape velocity (known as momentum-regulated feedback). In this case, we compare the momentum released in supernovae (or other feedback mechanisms, like radiation pressure) to the momentum required to lift the gas out of the halo at the escape velocity.\footnote{Note that requiring the feedback to eject gas from the halo is a stringent criterion -- gas could instead be ejected into the circumgalactic medium. Allowing for this additional effect will only increase the effects of bursts (see \S \ref{other-bursts}).}   Then,
\begin{equation}
\eta_p = \epsilon_p \pi_{\rm fid} \left( {10^{11.5} \, M_\odot \over m_h } \right)^{1/3} \left( {9 \over 1+z} \right)^{1/2},
\label{eq:etap_fiducial}
\end{equation}
where $\pi_{\rm fid}$ is the momentum injected per supernova in units of $2 \times 10^{33}   \ \rm{g \, cm/s}^2$ (which equals the momentum input from a Salpeter IMF with solar metallicity) and a fraction $\epsilon_p$ of this momentum is used to drive a wind.

\citet{furl17-gal} found that reasonable fits to the luminosity functions at $6 < z < 8$ could be obtained with $\epsilon_p \approx 5$.\footnote{That model also required imposing a maximum star formation efficiency on massive galaxies to prevent overproducing bright sources. Bursts are most significant in very small, early galaxies, so we ignore that complication here.} We take this as our fiducial value here.

This feedback scenario implicitly assumes that much of the energy injected by supernovae is lost to radiative cooling. If that energy is available, feedback can be stronger, especially in smaller galaxies. \citet{furl17-gal} showed that such models could also provide reasonable fits to the luminosity function. We do not show any such  energy-regulated scenarios here, but we note that they have qualitatively similar behavior in the bursty phase as momentum-regulated models. 

\subsection{Bursts in turbulent disc galaxies} \label{disc-burst}

While the approach described above has the merits of simplicity, it lacks some physical context that could affect the burst cycles. In particular, the model so far treats $f_{\rm sf}$ as a free parameter, so that all the gas is eligible for star formation, and we prescribe the outflow strength. We now consider a model in which these galaxies are treated as star-forming discs. This provides a way to estimate both of these parameters in a more self-consistent fashion. This approach also introduces a threshold to star formation, as discussed in the context of the toy model in section \ref{toy-threshold}. Of course, these early galaxies may not form stable discs, but a qualitatively similar burst history will likely hold for turbulent gas clouds.

To this end, we assume that accreted gas settles into a rotationally-supported disc.  We assume that turbulence  supports the disc in the vertical direction. Then, the surface density is $\Sigma_g = 2 h \bar{\rho}$, where $\bar{\rho}$ is the disc density and $h$ is the scale height, given by
\bq
h \approx {c_{\rm eff} \over \sqrt{\phi} \Omega},
\eq
where $c_{\rm eff}$ is the effective velocity dispersion (here assumed turbulent), $\phi$ is of order unity for a thin self-gravitating disc \citep{thompson05, munoz12}, and $\Omega$ is the angular velocity. 

We then assume that gravitational instability -- which triggers fragmentation to high densities, and so is ultimately responsible for star formation -- is determined by the Toomre criterion \citep{toomre64}. Following \citet{orr19}, we allow for the presence of stars in the disc by taking an effective Toomre parameter
\bq
\tilde{Q}_{\rm gas} = {\sqrt{2} \sigma_R \Omega \over \pi G \Sigma_t},
\eq
where $\Sigma_t$ is the total surface density, $\sigma_g$ is the velocity dispersion (typically assumed to be turbulent), and for isotropic turbulence the radial velocity dispersion is $\sigma_R = \sigma_g/\sqrt{3}$, . Gravitational fragmentation sets in when $\tilde{Q}_{\rm gas} \sim 1$. Intuitively, accretion increases the surface density $\Sigma_g$ until the massive disc undergoes fragmentation and star formation begins. Feedback from the star formation stirs turbulence and evacuates gas until $\tilde{Q}_{\rm gas} \approx 1$. Assuming the star formation and feedback can come into equilibrium, the system will reach a state of marginal stability in which the disc evolves along a sequence of such states \citep{faucher13,krumholz18}. \citet{orr19} examined this in particular, finding that a feedback delay induces oscillations to the star formation rate around the quasi-equilibrium solution, but they considered systems in which the delay was short compared to other timescales, which will not be the case at high redshifts.

Because the instability threshold depends upon the gas velocity dispersion, we must track that quantity as well.\footnote{Tracking $\sigma_g$ is not necessary under the quasi-equilibrium assumption, because it is fixed by the instantaneous star formation rate.} We therefore add a new equation to the system describing the galaxy evolution \citep{orr19},
\bqa
\dot{\sigma}_{g} & = & \dot{\sigma}_{\rm SN}^d - {\sigma_g \over t_{\rm eddy}} - \dot{\sigma}_w - \dot{\sigma}_\star + f_v v_c { \dot{m}_{c,g} \over m_g }. \label{eq:sig-evol}
\eqa
Here $\dot{\sigma}_{\rm SN}^d$ is the rate at which supernovae inject momentum, $\sigma/t_{\rm eddy}$ is the rate at which turbulent eddies dissipate it, $\dot{\sigma}_w$ is the rate at which winds carry away the momentum, $\dot{\sigma}_\star$ is the rate at which momentum is dissipated by star forming gas, and the last term is the rate at which accretion generates turbulence in the disc. 

The terms on the right side of equation~(\ref{eq:sig-evol}) all have simple physical interpretations. The term $\dot{\sigma}_\star$ represents the momentum dissipated during the star formation process. We assume that the turbulent velocity is fully mixed with the star-forming material, so this is simply $\dot{m}_\star/m_g$.  The second term is the rate at which the turbulent cascade dissipates energy. Assuming that the eddies are isotropic, their maximum size is the disc thickness $h$, and the turnover timescale is just $t_{\rm eddy} \sim \Omega^{-1}$. The last term is the momentum advected by accretion; we assume that the typical turbulent velocity dispersion of this material is a fraction $f_v$ of the halo circular velocity $v_c$. We set $f_v=0.1$ for concreteness, but its choice does not significantly affect our results. 

Finally, the two remaining terms account for the momentum injected into the ISM by star formation. Because we allow  feedback to expel gas from the galaxy -- and hence carry away momentum -- we write 
\bq
\dot{\sigma}_{\rm SN}^d -  \dot{\sigma}_w = (1 - f_{\rm out}) \VEV{ P_\star / m_\star } \dot{m}_\star^d/m_g,
\label{eq:sigma-evol}
\eq
where $f_{\rm out}$ is the fraction of the ISM material ejected by the feedback (see below) and $\VEV{P /m_\star}$ is the momentum injected per unit star formation. Note that we use the time-delayed effective star formation rate $\dot{m}_\star^d$ to compute the injection rate. This expression assumes that the momentum is injected uniformly throughout the ISM. We use the estimate for $\VEV{P_\star/m_\star}$ from \citet{furl20-disc} (which is taken from \citealt{martizzi15}), multiplied by a factor $0.75$ in order to match our fiducial simplified feedback model.\footnote{Specifically, we scale the feedback strength so that the total stellar mass density at $z=6$ is identical in our disc model and fiducial model, as described in section \ref{sf-history}.}

One advantage of the disc approach is that it provides a more physically-motivated picture for the fraction of the gas reservoir available for star formation, $f_{\rm sf}$. In the local Universe, this corresponds to the fraction of gas in molecular clouds. Rather than try to follow the chemistry of these clouds, we take a simpler approach by assuming that star formation is driven by gravitational instability in a turbulent medium, so $f_{\rm sf}$ is a function of the Toomre parameter. Motivated by simulations of turbulent discs, \citet{orr19} take
\bq
f_{\rm sf} = 0.3 (2/ \tilde{Q}_{\rm gas})^{\beta},
\label{eq:fsf-orr}
\eq
where 
\bq
\beta = -2 \log \left( {\Sigma_g \over M_\odot \ {\rm pc}^{-2}} \right) + 6
\eq
and the constants are chosen as in \citet{orr19}. Because our systems are so small (unlike \citealt{orr19}), we do not impose a maximum on $f_{\rm sf}$, except that it cannot exceed unity.  This form has not been tested in systems like the tiny galaxies we are interested in here; fortunately, the results are not particularly sensitive to the details of the $f_{\rm sf}$ prescription.

In this picture, where the injected momentum is used to support the disc, it is also natural to use a turbulent disc model to estimate the fraction of the gas that is unbound through feedback, providing a more physically-motivated way to estimate $\eta$. We use the model of \citet{hayward17} (see also \citealt{thompson16}), in which momentum is injected uniformly into a turbulent disc, and in which gas elements are ejected if they can be accelerated to the escape velocity within one eddy turnover time. This condition (which also depends on the momentum injected per unit mass of star formation, $\VEV{P_\star /m_\star}$) determines the fraction $f_{\rm out}$ of  gas that is ejected. The resulting mass-loading parameter is very close to a power law, $\eta \propto m_h^{-1/3}$, as expected for momentum-regulated feedback \citep{furl20-disc}.  

Finally, we note that this model is not fully self-consistent, as the condition for gas to escape the disc relies on assumptions about the turbulent density distribution -- information which is not contained in equation~(\ref{eq:sigma-evol}). It is therefore possible for $\sigma_g$ to increase beyond the orbital velocity (and even the escape velocity). To avoid unphysical effects, we limit $\sigma_g < v_c(r_{1/2})$, where $v_c(r_{1/2})$ is the circular velocity at the disc half-mass radius. In practice, this only happens when feedback has already ejected all but a tiny fraction of the gas, and the treatment of this limit does not significantly affect our results.

\section{The evolution of bursty galaxies} \label{ind-gal}

In this section, we consider how individual galaxies evolve in the presence of delayed feedback. 

\begin{figure*}
	\includegraphics[width=\columnwidth]{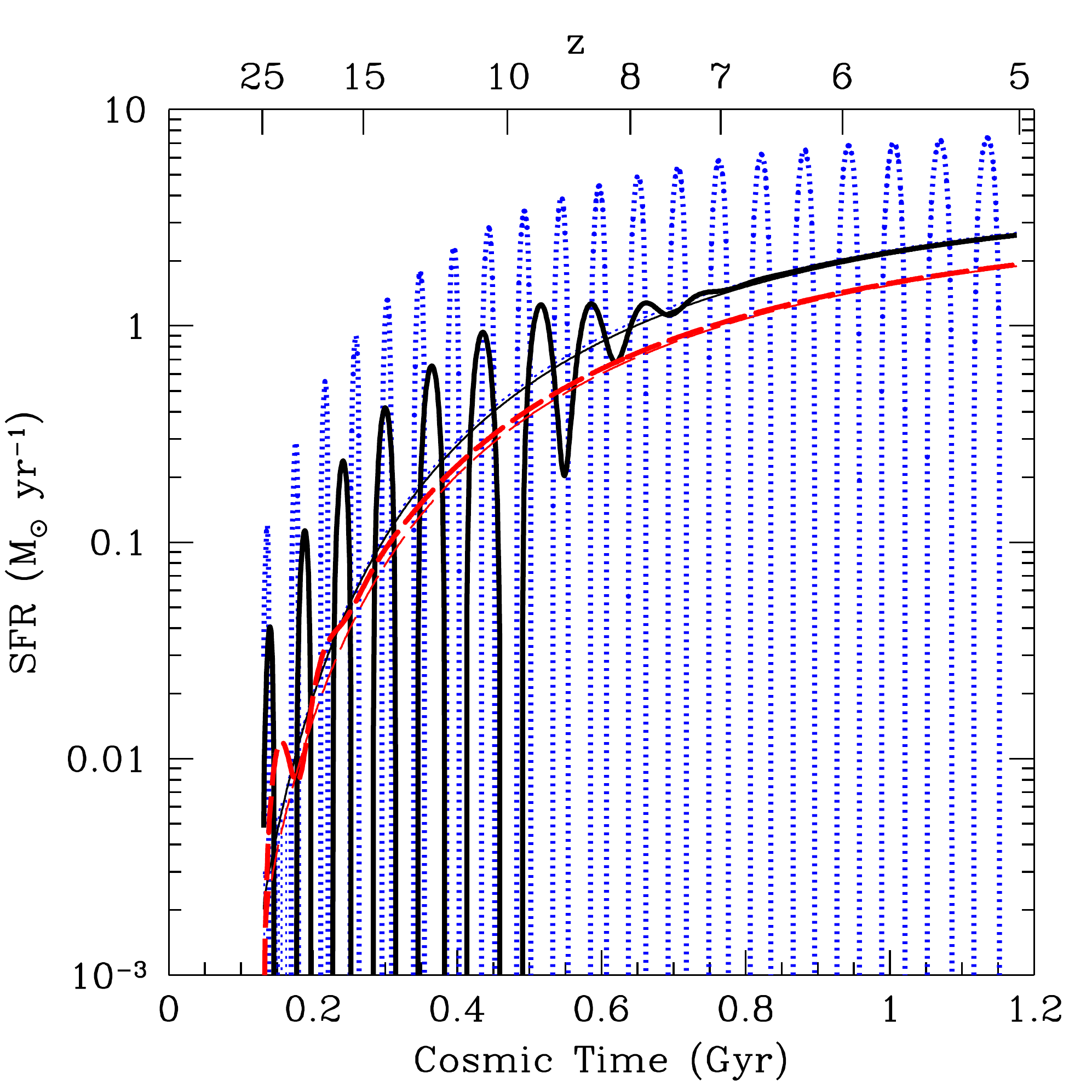} \includegraphics[width=\columnwidth]{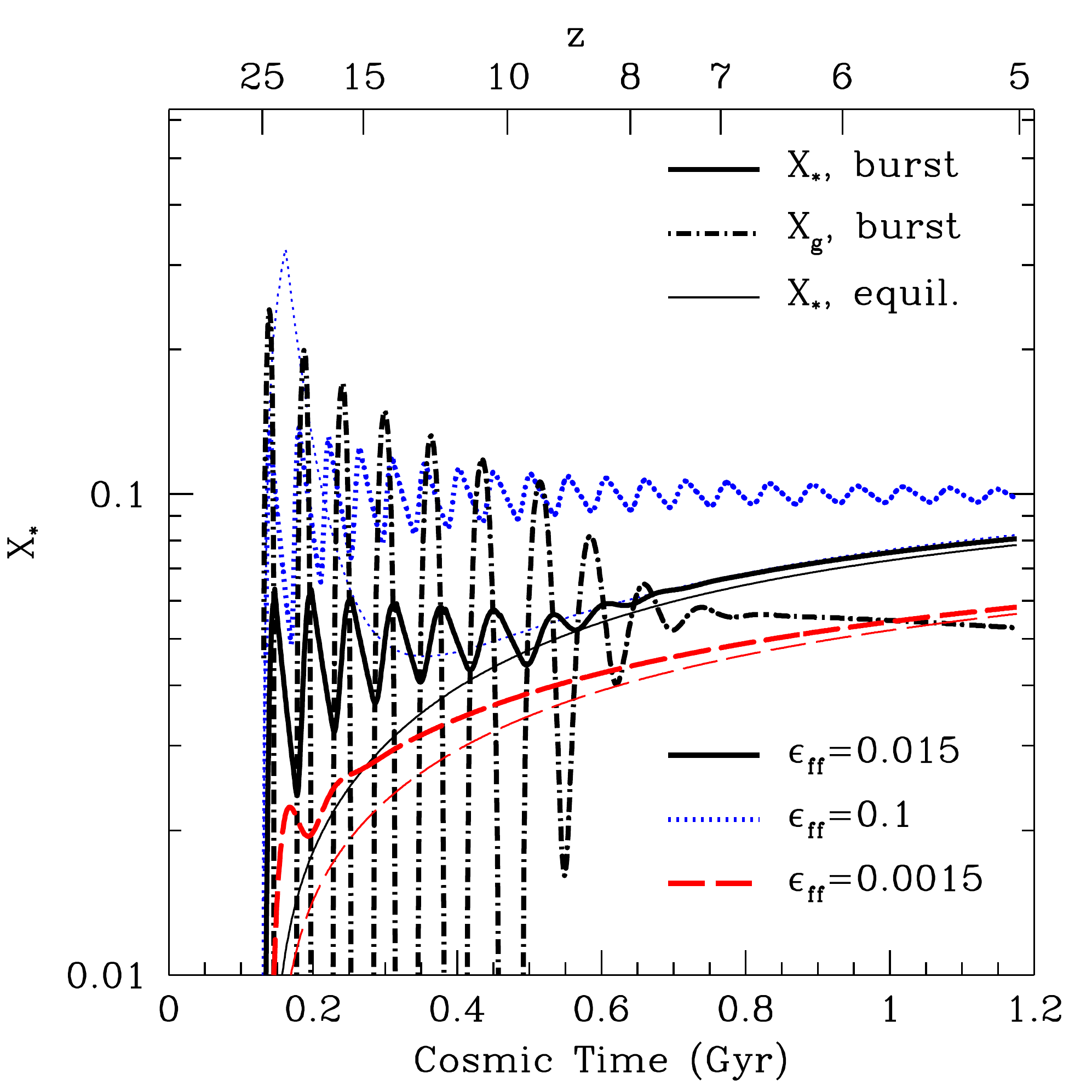} 
    \caption{Example star formation histories in our basic burst model. All curves follow a halo that begins forming stars at $z=25$, which has $m_h \approx 10^{11} \ M_\odot$ at $z=5$, and assume momentum-regulated feedback with $\epsilon_p=5$ injected over a delay interval of 5--30~Myr. The thick curves use our bursty model, with $\epsilon_{\rm ff}=0.1,\,0.015$, and 0.0015 (dotted, solid, and dashed, respectively). The left panel shows the star formation rate, while the right panel shows the fraction of $X_\star = m_\star/(\Omega_b m_h/\Omega_0)$
    baryons turned into stars. The thin curves take the same parameters but assume no delay between star formation and feedback. In the right panel, the dot-dashed curve shows the fraction of 
    baryons in the gas reservoir in the bursty model with $\epsilon_{\rm ff}=0.015$.}
    \label{fig:burst-basic}
\end{figure*}

\subsection{Bursts and the basic parameters of star formation and bursts}

In Figure~\ref{fig:burst-basic} we show some example star formation histories using our simple model of delay-driven bursts. In each case, we consider a halo that begins forming stars at $z=25$, which grows to $m_h \approx 10^{11} \ M_\odot$ at $z=5$. The thick curves use our fiducial model with momentum-regulated feedback and $\epsilon_p=5$ injected at a constant rate over a delay interval $5$--30~Myr. The dotted, solid, and dashed curves take $\epsilon_{\rm ff} = 0.1,\,0.015$, and 0.0015. For comparison, we also show the same models but with instantaneous feedback (thin curves).

The left panel shows that a delay in the feedback injection can result in a strongly fluctuating star formation rate, with the amplitude of the oscillations depending on the star formation efficiency: for a very low value of $\epsilon_{\rm ff}$, the solution quickly converges to the quasi-equilibrium case, but for efficient star formation the fluctuations remain strong throughout this period. In the intermediate case (which uses approximately the observed efficiency of local molecular clouds), the oscillations are strong at first but damp out as the galaxy grows.

The right panel shows the fraction of the halo's baryons that are processed into stars. Without a feedback delay, the fraction $X_\star \equiv m_\star/m_a$, where $m_a = (\Omega_b/\Omega_0) m_h$,  slowly and smoothly increases as the halo grows in mass (and becomes more resistant to feedback, with $\eta \propto m_h^{-1/3}$ in our model).\footnote{Note that in \citet{furl20-disc} we showed that $X_\star$ is largely independent of the star formation efficiency. However, if $\epsilon_{\rm ff}$ is very small, there is some small dependence, which we see here.} But with the feedback delay included, we see results similar to the toy models in Figure~\ref{fig:toy-models}, where $X_\star$ initially oscillates around a roughly constant value, well above the expectation for the quasi-equilibrium solution.\footnote{Note that the $\epsilon_{\rm ff}=0.1$ instantaneous feedback model also has a strong transient feature following the initial turn-on. In this regime, star formation initially outpaces feedback because it occurs so rapidly.} 

The toy model helps illuminate these results. Recall that model predicted that the stellar mass fraction oscillated around $X_\star^{\rm net} \approx \epsilon_{\rm ff} u_{\rm SN}/4$. For the fiducial model here, that is $X_\star \approx 0.03$ at the initial time, which is quite close to the actual value. 
Importantly, we see that this burst value does depend on $\epsilon_{\rm ff}$, although it is somewhat less sensitive than predicted by the toy model (here, $X_\star \propto \epsilon_{\rm ff}^{0.4}$ in the initial phases). We emphasize that this is in strong contrast to the quasi-equilibrium solutions, in which the efficiency is nearly independent of $\epsilon_{\rm ff}$. 

These galaxy models do show one important feature that is not reflected in the toy model, the transition from the burst phase to the quasi-equilibrium evolution as the halo grows. We can, however, understand this transition with reference to that model. Physically, this occurs because  the galaxy grows massive enough that feedback can no longer eject all the galaxy's gas (again, in our models $\eta \propto m_h^{-1/3}$).  The transition to smooth star formation will begin when a burst of stars can no longer evacuate the gas reservoir, or when $X_\star^{\rm net} \eta \sim 1$. Because $X_\star^{\rm eq} \sim 1/\eta$ in the quasi-equilibrium regime, this is equivalent to the condition that $X_\star^{\rm net} \sim X_\star^{\rm eq}$, or in other words the $X_\star$ curves should oscillate until they meet the quasi-equilibrium case. We find that our models do indeed have $X_\star^{\rm net} \eta \sim 1$ when the transition occurs, as expected. In other words, the burst models only settle onto the quasi-equilibrium curve when the halo becomes tightly bound enough to retain gas between star formation cycles (see the dot-dashed curve, which shows the gas fraction $X_g$ for the $\epsilon_{\rm ff}=0.015$); this does not happen for the $\epsilon_{\rm ff}=0.1$ case because the bursts are so strong.

\begin{figure*}
	\includegraphics[width=\columnwidth]{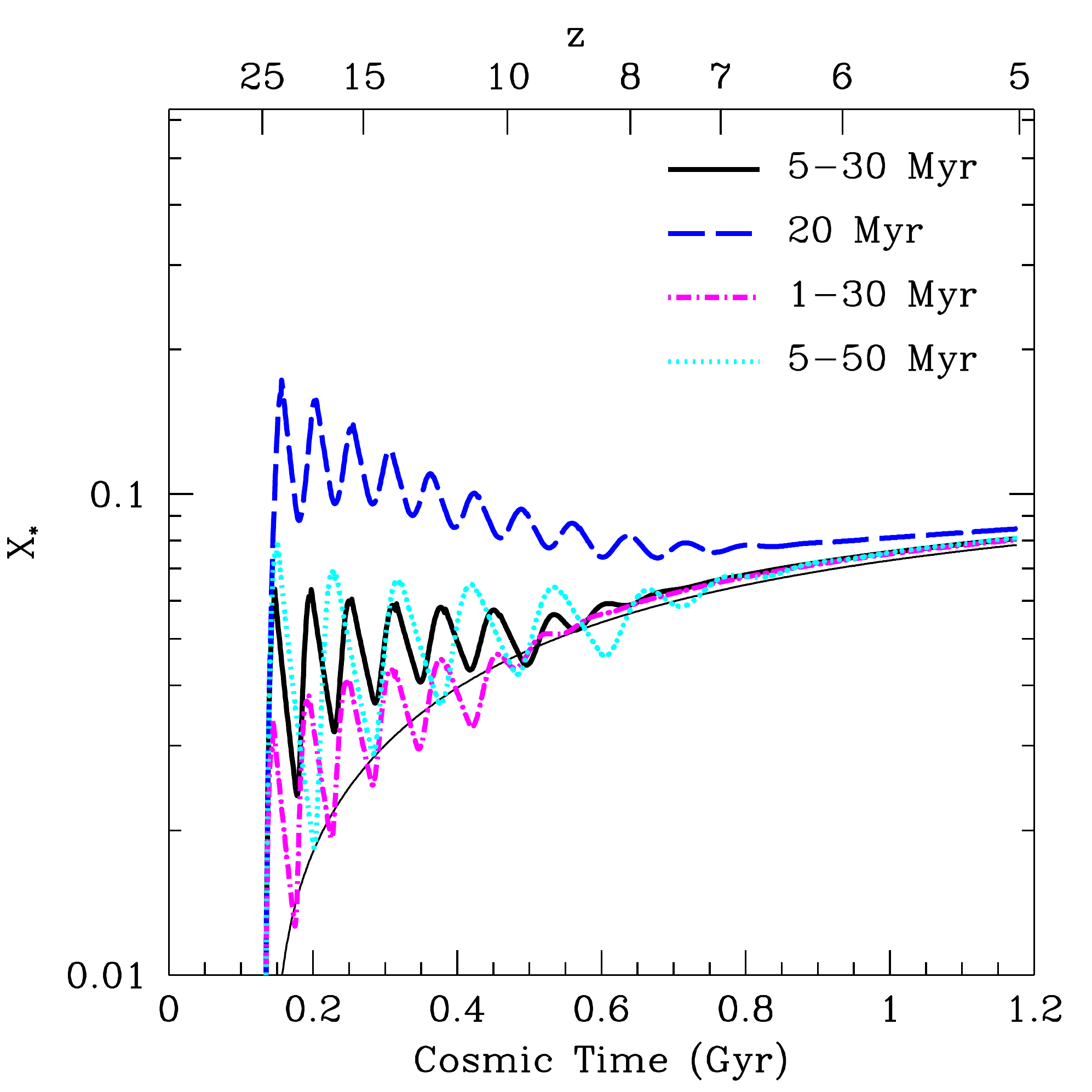} \includegraphics[width=\columnwidth]{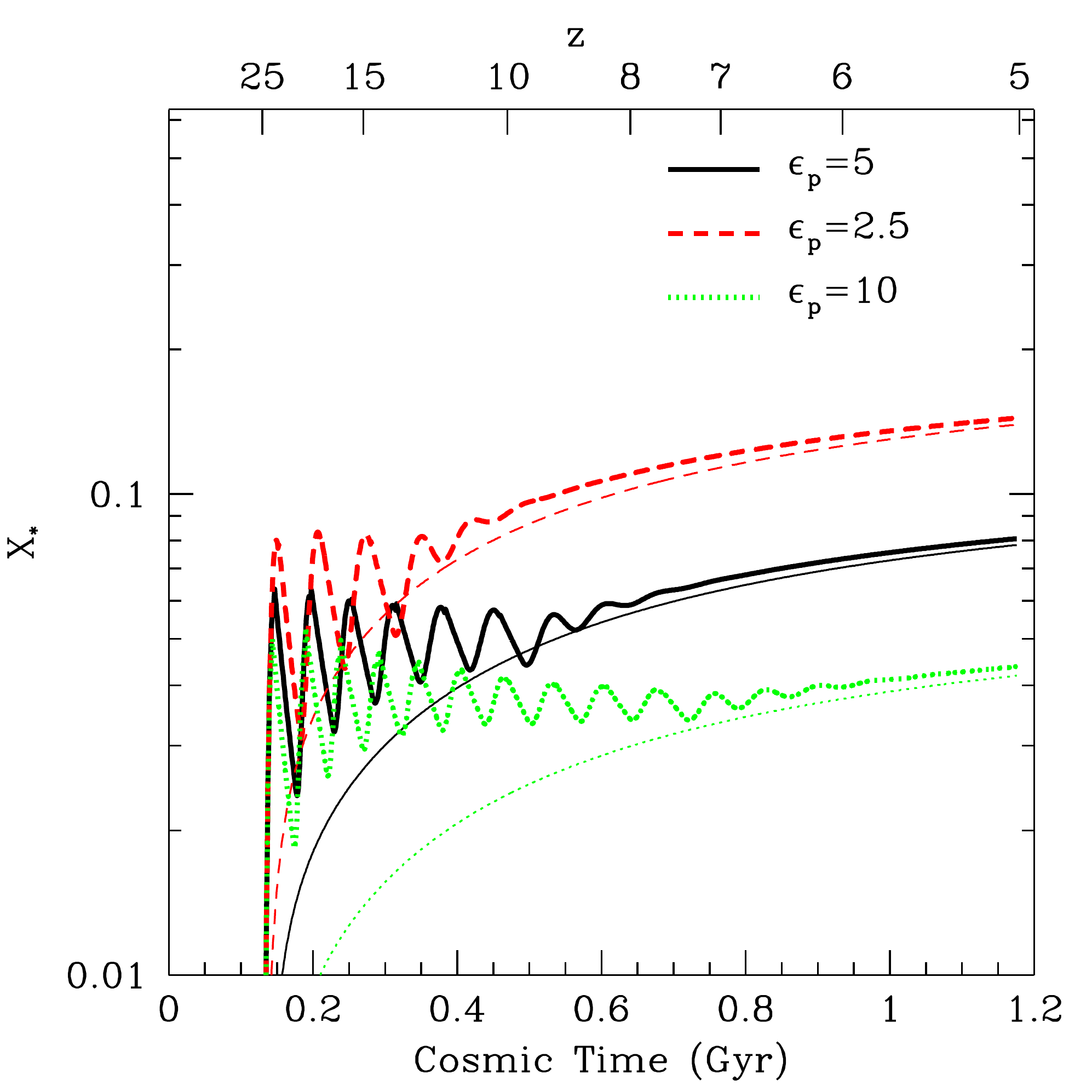} 
    \caption{The effects of the feedback delay (left panel) and amplitude (right panel) on the fraction of halo gas converted into stars, $X_\star$. All curves follow a halo that begins forming stars at $z=25$, which has $m_h \approx 10^{11} \ M_\odot$ at $z=5$, and assume $\epsilon_{\rm ff}=0.015$. In the left panel, all curves assume momentum-regulated feedback with $\epsilon_p=5$ but vary the interval over which that energy is injected. (The dashed curve assumes it is injected instantaneously after a 20~Myr delay). In the right panel, all curves take a feedback delay of 5--30~Myr but assume  $\epsilon_p=2.5,\,5$, and 10 (dotted, solid, and dashed, respectively).     }
    \label{fig:burst-basic-uSNb}
\end{figure*}

\begin{figure*}
	\includegraphics[width=\columnwidth]{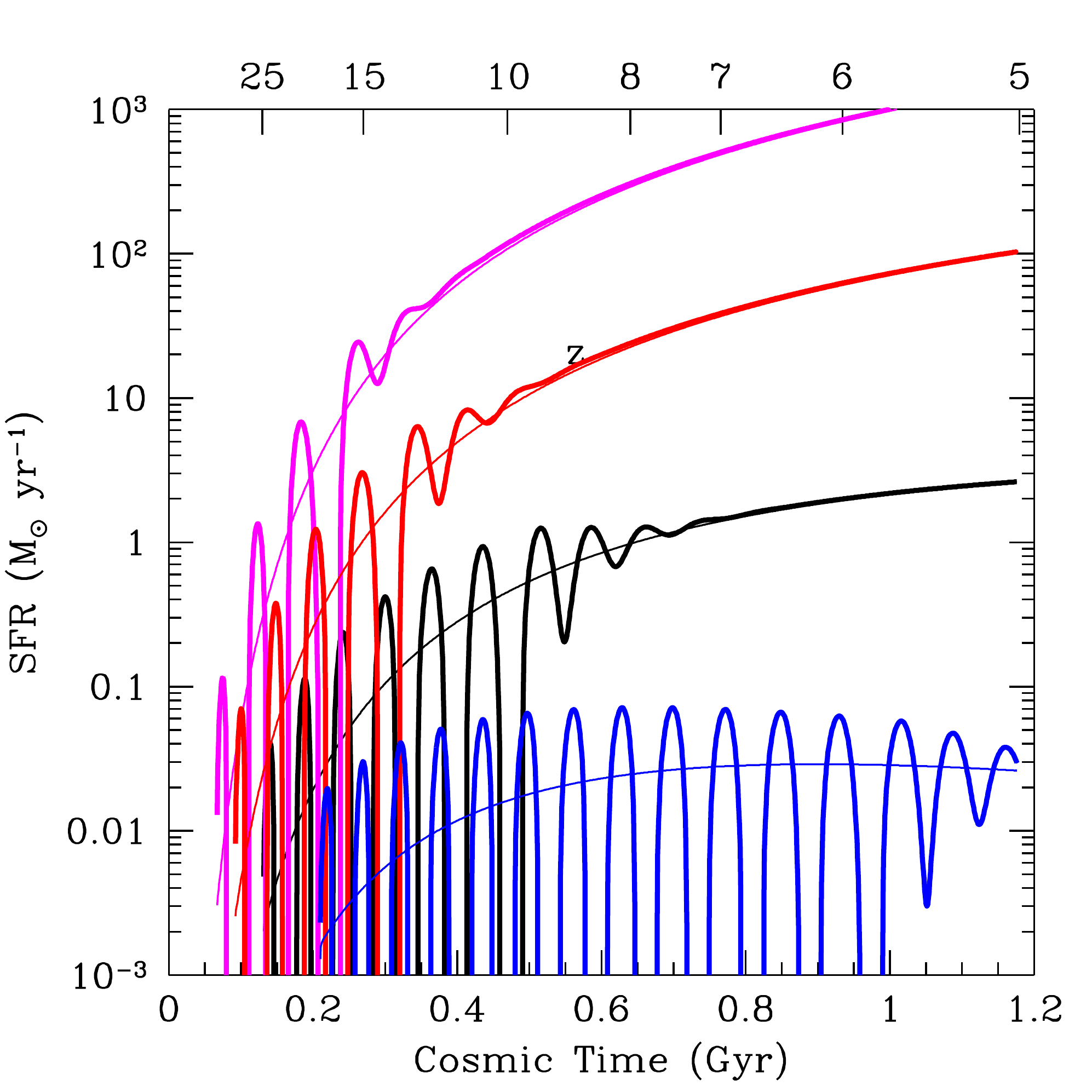} \includegraphics[width=\columnwidth]{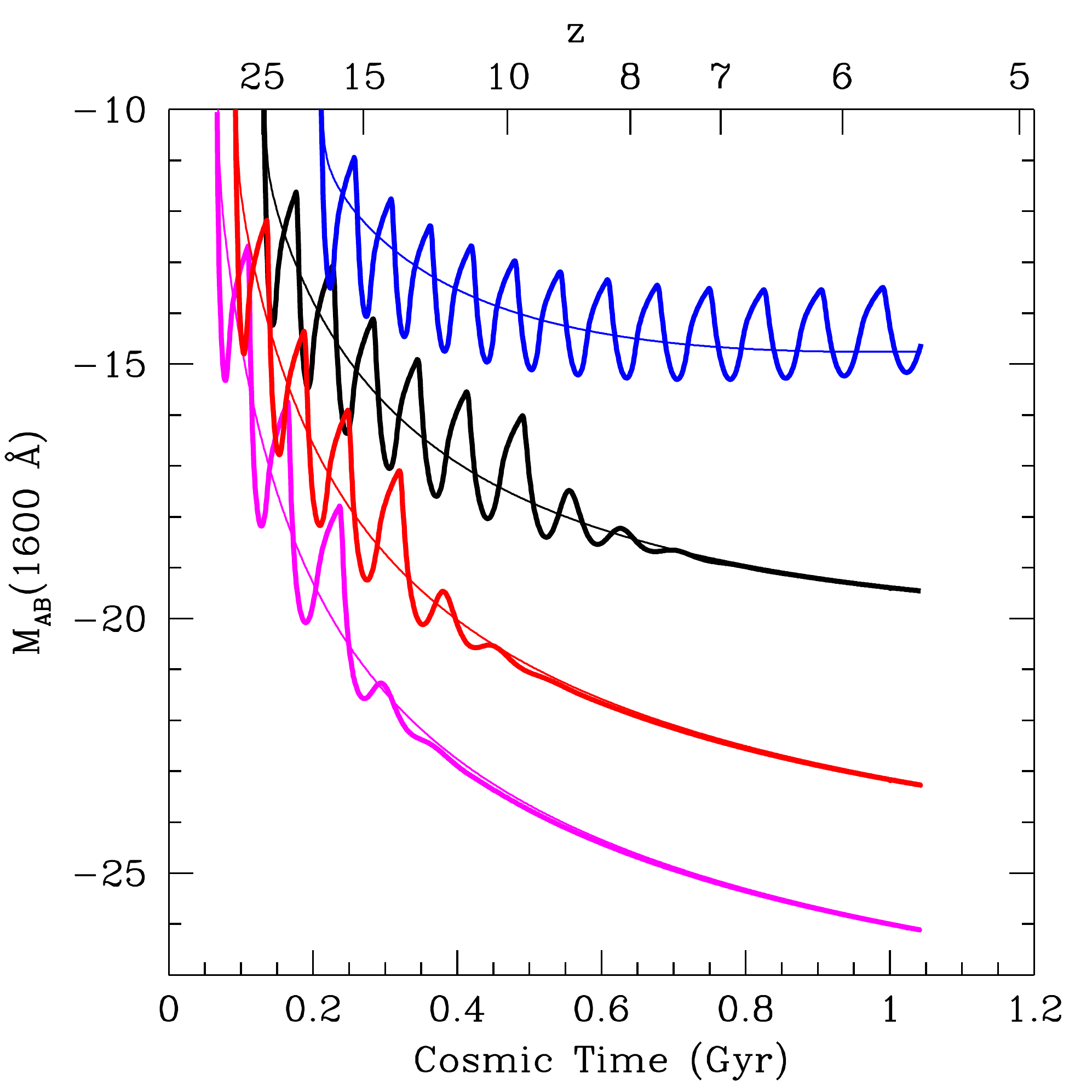} 
    \caption{\emph{Left:} Identical to the left panel of Fig.~\ref{fig:burst-basic}, but varying the halo masses. From bottom to top, the curves begin star formation at $z=18,\,25,\,32$, and 40, respectively, which correspond to halo masses at $z=5$ of $m_h \approx (0.04,\,1,\, 14, \, 120) \times 10^{11} \ M_\odot$, respectively. All curves use our fiducial parameters. Thick curves use the burst model; thin do not. \emph{Right:} Corresponding luminosities, computed with the \textsc{ares} code.}
    \label{fig:burst-basic-z}
\end{figure*}

In Figure~\ref{fig:burst-basic-uSNb}, we explore how this burst phase depends on the feedback assumptions. The left panel holds the feedback amplitude constant but varies the time interval over which it is injected. Unsurprisingly, we find that the key parameter is the \emph{initial} delay time: allowing the feedback to begin 1~Myr after star formation (dot-dashed curve) does not entirely remove the burstiness, but it does decrease the amplitude of the ``overshoot," so that the bursts do not depart strongly from the quasi-equilibrium curve. However, setting the delay to 20~Myr increases that overshoot significantly: as expected from the toy model, $X_\star^{\rm net} \propto \epsilon_{\rm ff} u_{\rm SN}$, with $u_{\rm SN}$ best represented by the minimum feedback time. In all of these curves, we have taken $\epsilon_{\rm ff}=0.015$ to match local measurements; as we have already seen, increasing that efficiency will also increase the burstiness, even if the delay time is small.

The right panel of Figure~\ref{fig:burst-basic-uSNb} holds the delay timescale constant but varies the overall strength of the feedback. This has a strong effect on the quasi-equilibrium solution but relatively little effect on the burst phase. As expected from the toy model, star formation during burst phases proceeds mostly before feedback becomes effective, so $X_\star$ is only modestly dependent on the feedback strength. 

\begin{figure*}
	\includegraphics[width=\columnwidth]{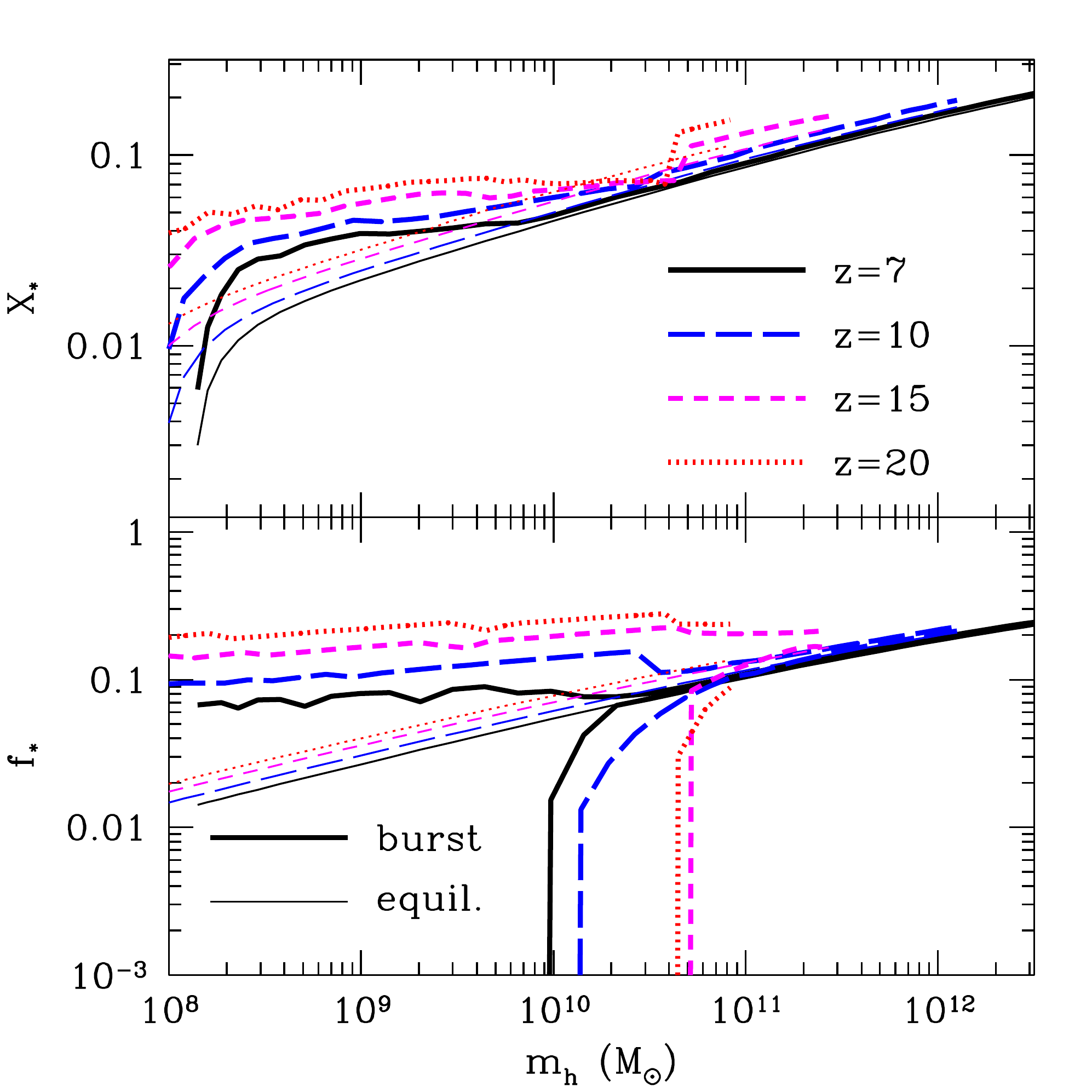} \includegraphics[width=\columnwidth]{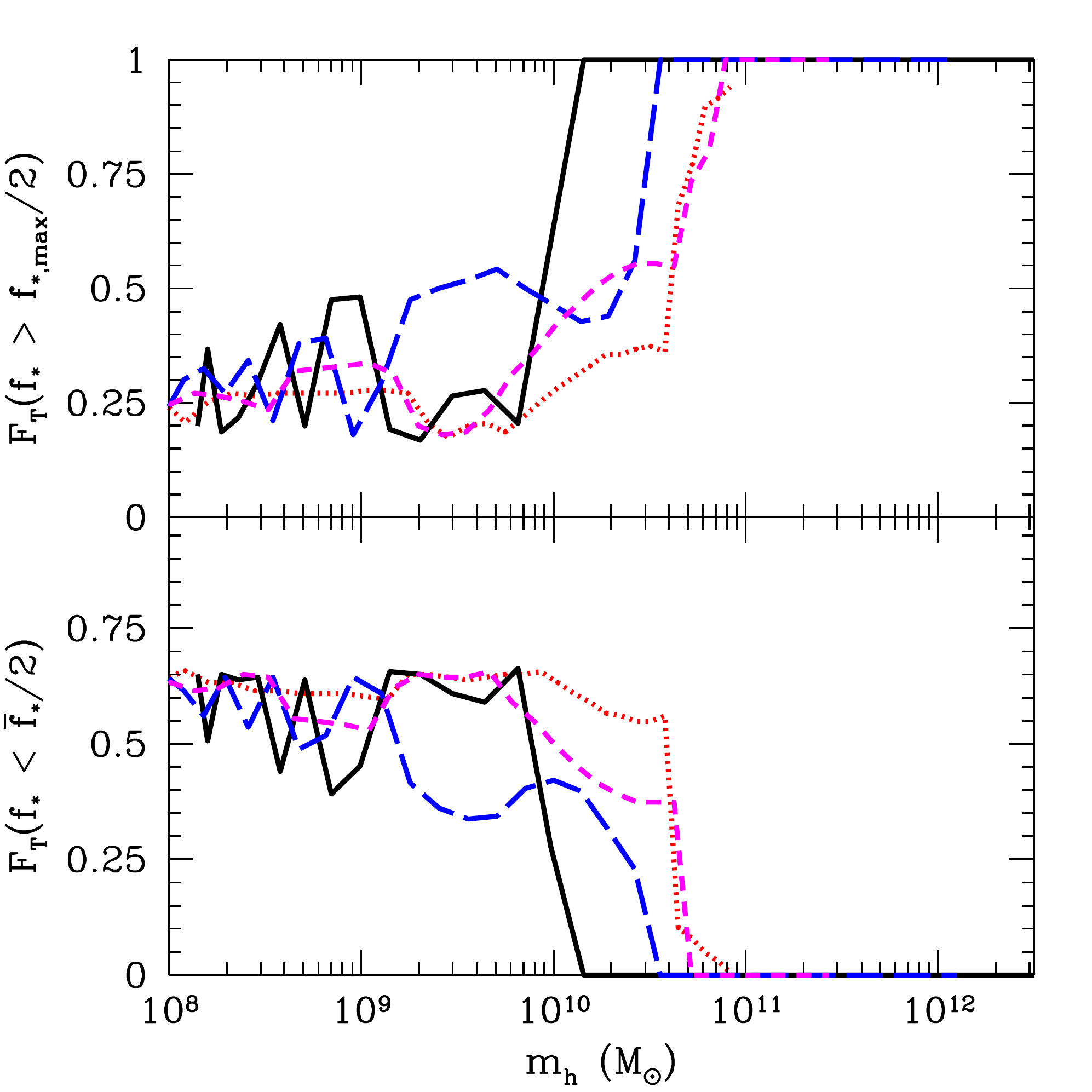} 
    \caption{Trends with halo mass in burst models across several redshifts. The thick curves use our fiducial parameters (momentum-regulated feedback with $\epsilon_p=5$ injected over a delay interval of 5--30~Myr and $\epsilon_{\rm ff}=0.015$). Thin curves ignore the feedback delay. \emph{Top left:} Time-integrated star formation efficiency as a function of mass. Note that, because the bursty models have strong time dependence, we take the average values over the previous 50~Myr here. \emph{Bottom left:} The minimum and maximum values of the instantaneous star formation efficiency, $f_\star$, over the previous 50~Myr. \emph{Top right:} The fraction of time during the previous 50~Myr for which $f_\star$ is greater than half its maximum value. \emph{Bottom right:} The fraction of time for which $f_\star$ is smaller than half its average value.}
    \label{fig:mass-trend}
\end{figure*}

Thus, our models show that high-redshift galaxies should transition between two star formation regimes: in small galaxies at very early times, bursts driven by the feedback delay dominate, and the time-integrated star formation efficiency depends reasonably strongly on the small-scale star formation efficiency $\epsilon_{\rm ff}$ and the timing of the feedback but only weakly on its strength. But once the halo is able to retain its gas supply, it transitions to the quasi-equilibrium solution in which $X_\star$ is nearly independent of $\epsilon_{\rm ff}$ and the feedback timing but depends strongly on the feedback amplitude. The transition occurs when $X_\star^{\rm net} \eta \sim 1$.

These qualitative conclusions are independent of other assumptions about stellar feedback (such as energy-regulated feedback) and the star formation prescription (as in our turbulent disc model). We do not show any such examples here but will explore those alternate models in section \ref{burst-pops}.

\subsection{Dependence on halo mass} \label{burst-mass}

Figure~\ref{fig:burst-basic-z} shows how the burstiness manifests in haloes with a range of masses. The haloes shown here begin forming stars at $z=18,\,25,\,32$, and 40, from bottom to top, which correspond to $z=5$ halo masses of $m_h \approx (0.04,\,1,\, 14, \, 120) \times 10^{11} \ M_\odot$, respectively.\footnote{Note that we do not include any kind of star formation quenching in our simple model; in reality, the most massive of these galaxies would likely be affected by virial shock heating \citep{faucher11} and/or AGN feedback, suppressing the enormous star formation rates seen here.} All of the curves use our fiducial parameter choices $(\epsilon_{\rm ff}=0.015$, $\epsilon_p=5$, and a 5--30~Myr delay). In most of these haloes, bursts are present during the initial phases of halo growth, only damping out once the halo becomes large (so that the high accretion rates and deep potential wells are able to overcome feedback from a single burst). 

In the right panel of Figure~\ref{fig:burst-basic-z}, we show how these star formation histories manifest in the observable UV luminosity. We use the \textsc{ares} code to synthesize the 1600~$\Angstrom$ luminosity of each object over its past star formation history \citep[as in][]{mirocha20-dust}, adopting the BPASS v1.0 single star models \citep{eldridge09} with a fixed metallicity of $Z=Z_{\odot}/5$ for simplicity. The crucial point is that the bursts induce a $\pm 1$ magnitude in the luminosities, or even larger at small masses. Even though the star formation rate can vary much more dramatically, the UV luminosity is sensitive to the star formation integrated over a timescale of $\ga 20$~Myr, the lifetime of very massive stars. Thus the effect on the luminosity is less extreme than one might naively expect. In section \ref{burst-lf}, we will discuss how this scatter affects the luminosity function of early galaxies.

\begin{figure*}
	\includegraphics[width=\columnwidth]{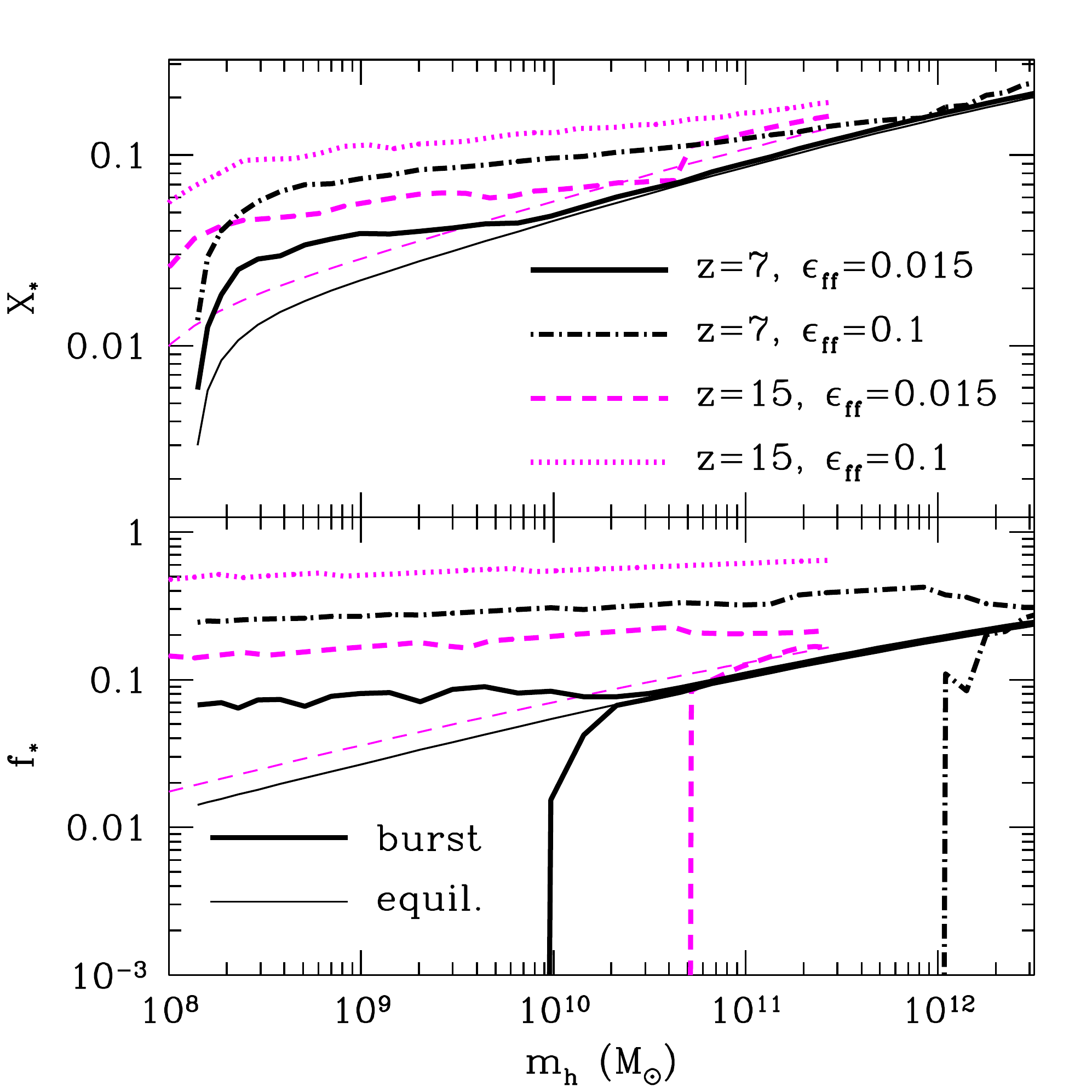} \includegraphics[width=\columnwidth]{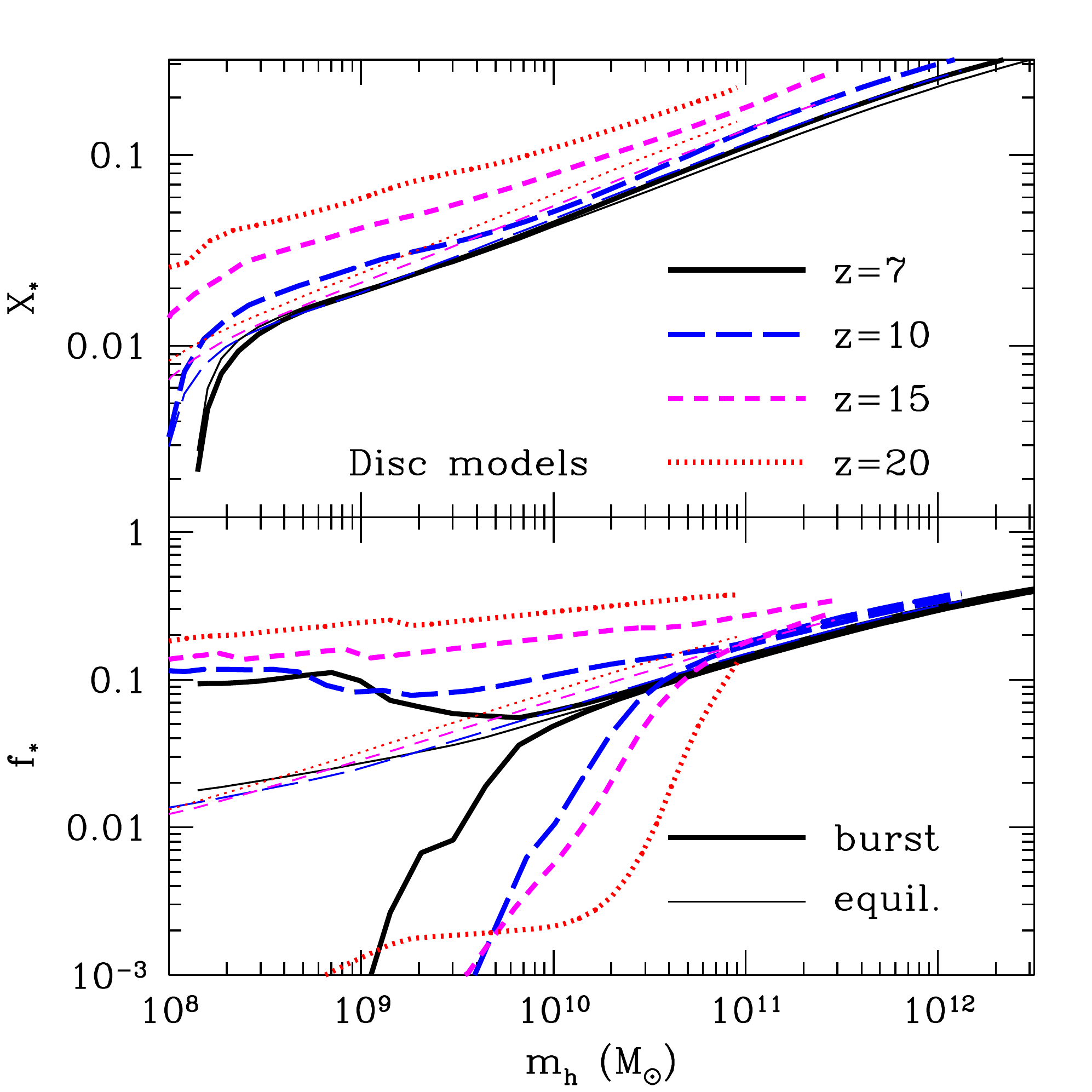} 
    \caption{As the left panels of Fig.~\ref{fig:mass-trend}, but comparing several different models. \emph{Left:} Models with different star formation efficiencies. The curves compare $\epsilon_{\rm ff}=0.015$ and 0.1 at $z=7$ and 15; the thick curves incorporate the feedback delay, while the thin curves do not. \emph{Right:} Identical to the left panels of Fig~\ref{fig:mass-trend}, but for the turbulent disc model.}
    \label{fig:mass-trend-mods}
\end{figure*}

\section{Populations of bursty galaxies} \label{burst-pops}

Now we shift our attention to consider the effects of feedback delay-driven bursts on the galaxy population as a whole. In Figure~\ref{fig:mass-trend}, we compare the star formation efficiencies and burstiness of models with a feedback delay (thick curves) with those with instantaneous feedback (thin curves). In the top left panel, we show the time-integrated star formation efficiency $X_\star$. Of course, in our models this can vary strongly with time during the bursty phase. We therefore average $X_\star$ over the previous 50~Myr. The choice of 50~Myr is arbitrary, and it leads to some artifacts in the curves (especially in the right panels). A different averaging period will certainly affect the details of these curves. But there is not a particularly well-motivated choice, unless one assumes one knows the feedback timescale well enough a priori.

The key point is that, when burstiness is significant, the integrated star formation efficiency $X_\star$ is elevated, because star formation overshoots the equilibrium solution. This overshoot depends only very weakly on the feedback amplitude $\eta$, which imprints the mass dependence on the quasi-equilibrium models ($X_\star \propto m_h^{1/3}$, as expected for momentum-regulated feedback). As a result, $X_\star$ has much weaker mass dependence in the bursty regime. This increases the star formation efficiencies of small halos, even without changing the strength of feedback, which has important implications for the star formation history of the Universe (see section \ref{sf-history}).

The bottom left shows another way of defining the star formation efficiency as an instantaneous quantity: $f_\star=\dot{m}_\star/\dot{m}_{c,g}$. Because $f_\star$ is a strong function of time in our bursty models, we show the minimum and maximum values over the previous 50~Myr.  The amplitude of the oscillations is quite large in the bursty phase: that regime corresponds to the complete evacuation of the gas reservoir (so that the star formation rate goes to zero), with only a narrow mass range (at any given redshift) over which the haloes transition to the quasi-equilibrium solution. Moreover, the overshoot is also apparent, with $f_\star$ exceeding the equilibrium solution by nearly an order of magnitude for the smallest haloes. 

The right panels show some more detail on the burst cycles. Here $F_T$ is the fraction of time (over the previous 50~Myr) for which the star formation rate is greater than half its maximum value (top panel) and the fraction that is is less than half its average value  (bottom panel).  In the bursty regime, the galaxies form stars rapidly for $\sim 25\%$ of the time in these models, spending slightly over half the time at very low star formation rates. 

Figure~\ref{fig:mass-trend-mods} shows that this behavior is generic across our suite of models. The left panels show how increasing the small-scale star formation efficiency $\epsilon_{\rm ff}$ affects the burstiness.  Recall that increasing this parameter increases the amount of star formation that occurs during each burst, which both increases $X_\star$ in that regime and enables the bursts to persist to higher masses. Note that the other qualitative effect of burstiness, flattening the $X_\star(m_h)$ curve, also occurs in this case. Although we do not show it explicitly, decreasing $\epsilon_{\rm ff}$ has the opposite effects.

The right panels of Figure~\ref{fig:mass-trend-mods} use the turbulent disc model described in section \ref{disc-burst}. The disc model slightly decreases the level of burstiness at high redshifts and has a smoother transition from strong bursts to the equilibrium solution, but the overall behavior is quite similar. (Note that we have fixed the feedback amplitude in this case so that $X_\star$ is similar at $z=7$ between the two models.) The disc model includes an additional condition for star formation (expressed through $f_{\rm sf}$, but that function is smooth so does not significantly affect the transition. It does result in a steeper mass dependence for $X_\star$ in the bursty regime, although it actually increases the overall overshoot somewhat.

\section{Discussion} \label{discussion}

\subsection{Burstiness and the luminosity function} \label{burst-lf}

\begin{figure}
	\includegraphics[width=\columnwidth]{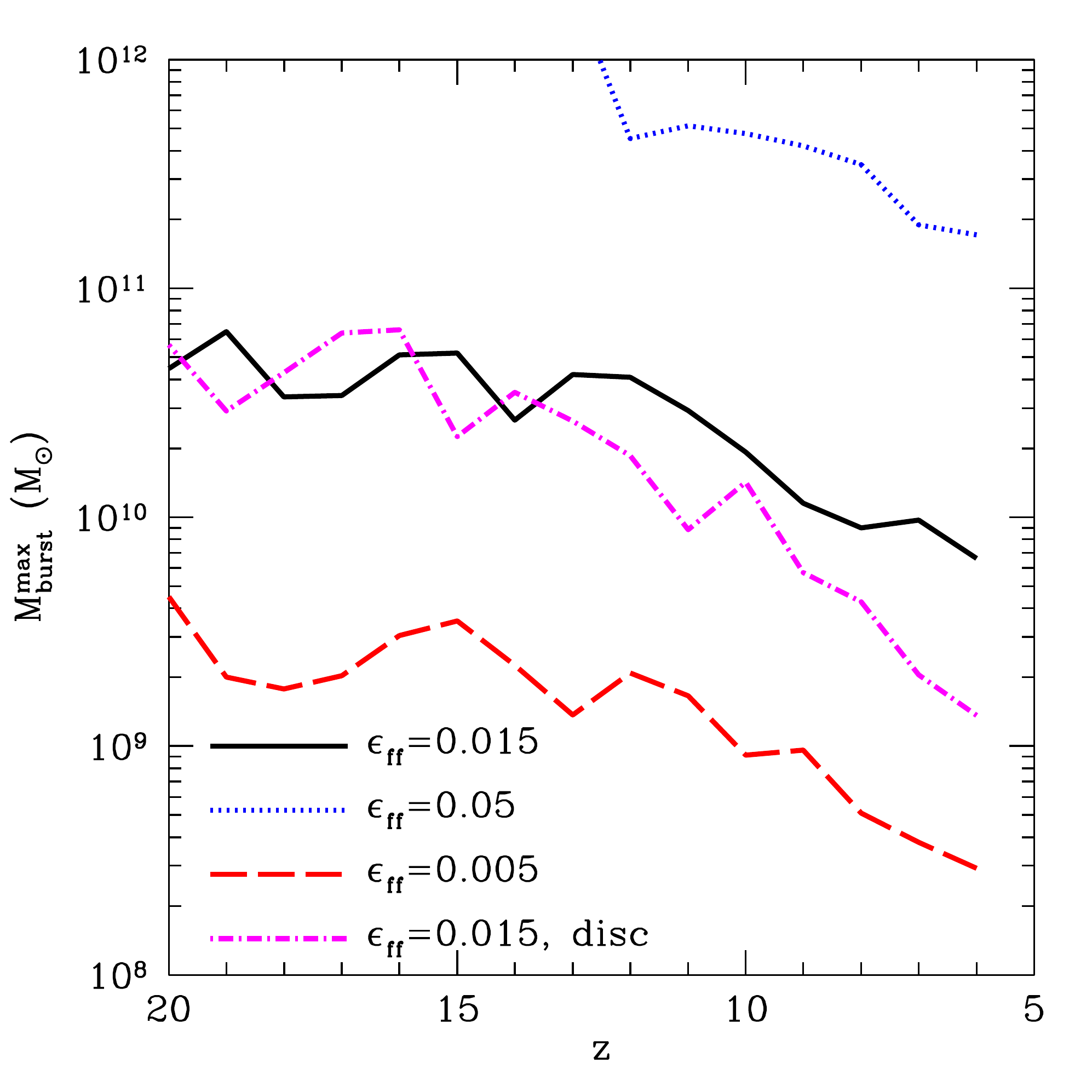} 
    \caption{ Threshold mass for ``steady" star formation. For $m_h < M^{\rm max}_{\rm burst}$, there is at least an order of magnitude variation in the instantaneous star formation efficiency over a 50~Myr time interval. The long-dashed, solid, and short-dashed curves takes $\epsilon_{\rm ff}=0.005,\,0.0015$, and 0.05, respectively, while the dot-dashed curve takes $\epsilon_{\rm ff}=0.0015$ for a turbulent disc model. if $\epsilon_{\rm ff}=0.1$, the threshold mass is $> 10^{12} \ M_\odot$ throughout this redshift interval.}
    \label{fig:mthreshold}
\end{figure}

We have seen in the previous sections that, when delays in the feedback induce burstiness, the star formation rate varies dramatically on short timescales, which induces variations of $\sim \pm 1$~mag in the luminosity as well. Figure~\ref{fig:mass-trend} shows that this burstiness sets in rather abruptly. Because the transition from the bursty regime to the quasi-equilibrium solution occurs quite rapidly, it is useful to define a threshold  $M^{\rm max}_{\rm burst}$ as the last mass at which the instantaneous, galaxy-averaged star formation efficiency $f_\star$ varies by at least an order of magnitude over the previous 50~Myr, for a given halo growth history. 
Figure~\ref{fig:mthreshold} shows this threshold mass at which this transition occurs for several models, as a function of redshift. 

We see that the threshold for burstiness depends strongly  on the assumed small-scale star formation efficiency parameter $\epsilon_{\rm ff}$: indeed, in this plot, we only vary $\epsilon_{\rm ff}$ by an order of magnitude, because for our usual choices of high/low efficiency ($\epsilon_{\rm ff}=0.0015$ and $\epsilon_{\rm ff}=0.1$), either no galaxies are bursty or all are. This will depend on other parameters of the model as well: if feedback is weaker, for example, the threshold mass decreases because the bursts have a harder time expelling the gas reservoir.

Nevertheless, the ``best guess" small-scale star formation efficiency, $\epsilon_{\rm ff}=0.015$, places the transition at $\sim 10^{10}$--$10^{10.5} \ M_\odot$ during reionization, which is comparable to the minimum halo mass observed so far in deep HST surveys at $z \sim 7$. 
Importantly, this implies that burstiness will preferentially affect faint galaxies: the properties of bright galaxies will remain nearly the same as the quasi-equilibrium case. But the excess star formation in faint galaxies (and its rapid time variability) does affect the overall galaxy population, as we explore in the next section.  

\begin{figure}
	\includegraphics[width=\columnwidth]{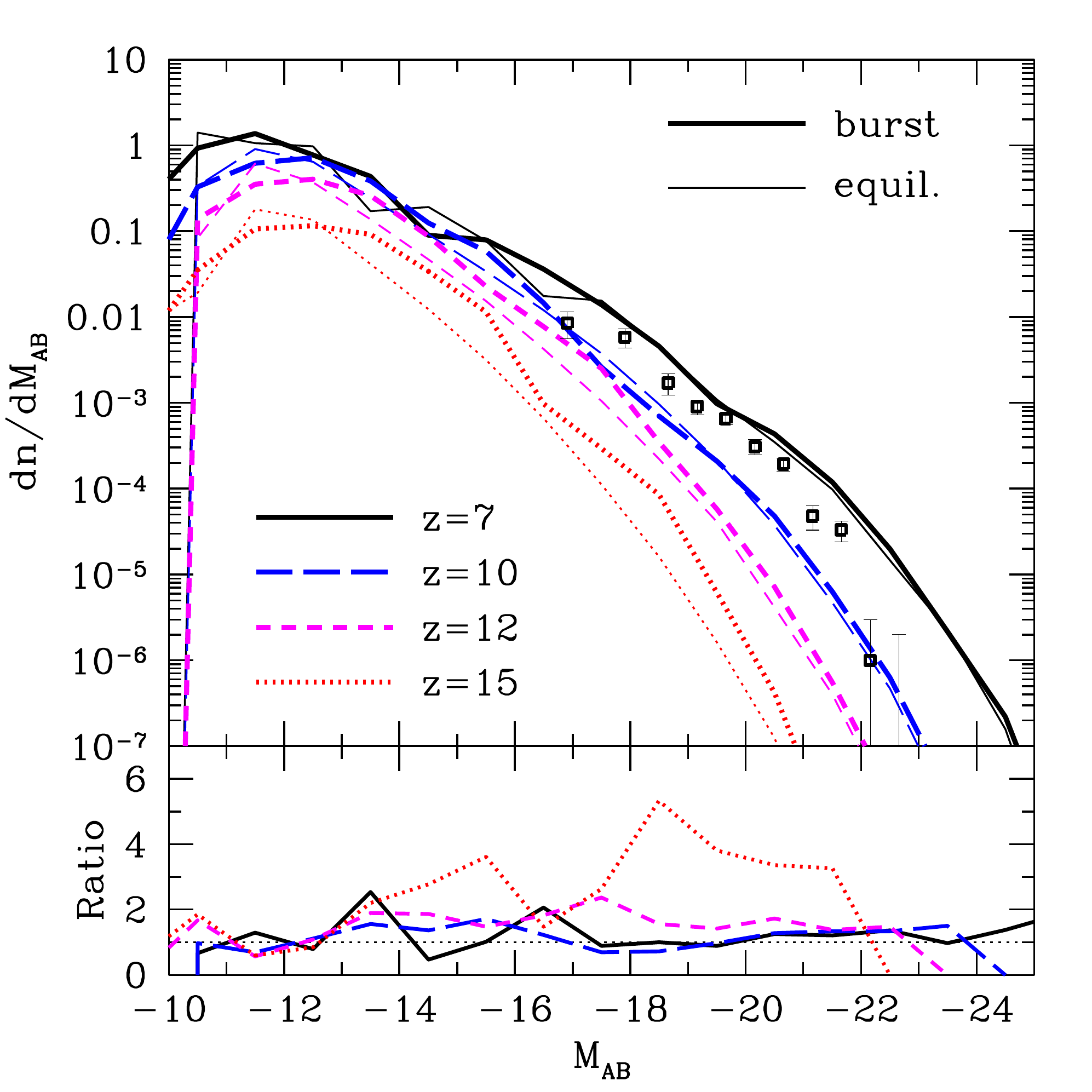} 
    \caption{The effect of delay-driven bursts on the observable galaxy luminosity function. In the upper panel, the thick and thin curves show the predicted luminosity function with and without a feedback delay, respectively, using our fiducial parameters ($\epsilon_{\rm ff}=0.015$, $\epsilon_p=5$, and a feedback delay of 5--30~Myr). We show results for $z=7,\,10,\,12$, and 15. The bottom panel shows the ratio of the predictions with bursts to those without them. The points with error bars show the luminosity function observed at $z=7$ by \citet{bouwens15} for context. }
    \label{fig:lf-effect}
\end{figure}

We explore this further in Figure~\ref{fig:lf-effect}, which shows how bursts affect the galaxy luminosity function in these models. To determine the luminosity function, we use the \textsc{ares} code to transform each halo's star formation history to a luminosity as a function of time (as described in section \ref{burst-mass}). We then use the \citet{trac15} halo mass function, randomly selecting a luminosity within 25~Myr of the assigned redshift according to our burst model, to assign haloes to luminosity bins. In the top panel, we show models incorporating delayed feedback with thick curves, while the thin curves assume instantaneous feedback; the bottom panel shows their ratio. For this purpose, we use our fiducial parameter set: $\epsilon_{\rm ff}=0.015$, $\epsilon_p=5$, and a feedback delay of 5--30~Myr.

The solid curves show the results at $z=7$, for which the bright end of the luminosity function is already fairly well-measured.  The points with error bars show the measurements from \citet{bouwens15}. In the observed range, the models with and without bursts are nearly identical, because these observations do not extend to very low-mass galaxies where bursts are still strong at $z \sim 7$. Both models do systematically overestimate the observed galaxy abundance in this range, but that can be remedied by adding a small amount of dust extinction, slowing accretion in massive galaxies (as in \citealt{faucher11}), or increasing the feedback amplitude by a factor $\sim 1.5$.

Overall, even in the faint regime delay-driven bursts have only a modest impact on the luminosity function at $z \la 10$. We would expect two qualitative effects. The first is the scatter in the luminosity induced by the burst cycles, as shown in Figure~\ref{fig:burst-basic-z}. This scatter is about $\pm 1$~mag at this redshift for low-mass halos, which broadens the luminosity distribution at the faint end somewhat. However, the luminosity function itself is fairly shallow in this regime, and the effect is quite modest. Thus it is not surprising that existing observations do not require bursts. 

The second effect is the overall enhancement in the average star formation rate due to the overshoot during each burst cycle. This is a modest effect at $z=7$, except in the smallest and faintest galaxies. But the enhancement is more pronounced at high redshifts (as shown in Fig.~\ref{fig:mass-trend}). The mass function is also much steeper at early times, and these two effects combine to shift the luminosity function systematically to higher luminosities at $z=15$ and especially $z=20$. In the latter case, the number of galaxies increases by a factor of a few in the observable range. 

Although delay-driven bursts do not appear to have a strong effect on the overall luminosity function, they can still have important effects when interpreting observations. For example, even the James Webb Space Telescope will only be able to observe the brightest galaxies at $z \sim 15$. If those galaxies are in the bursty regime, the luminosity of any one system can vary by several magnitudes on short timescales. This breaks the usual association between host mass and galaxy luminosity, and because the mass function is so steep at these redshifts, one is much more likely to find a relatively small galaxy near its peak luminosity than a massive system at its ``average" level. One must be very cautious in interpreting such systems. 

On the other hand, the enhancement to the luminosity function at this early time in the bursty mode does suggest that it can be probed even if the luminosity function can only be constrained to order unity. If the small-scale star formation efficiency is large, this can be a fairly dramatic effect. For example, if $\epsilon_{\rm ff}=0.1$, we find enhancements to the $z=12$ and $z=15$ luminosity functions by factors of five and ten (respectively) relative to the quasi-equilibrium model for $M_{AB} \la -14$. This more gradual evolution to the galaxy abundance would of course be good news for high-$z$ galaxy searches with forthcoming instruments.

\begin{figure}
	\includegraphics[width=\columnwidth]{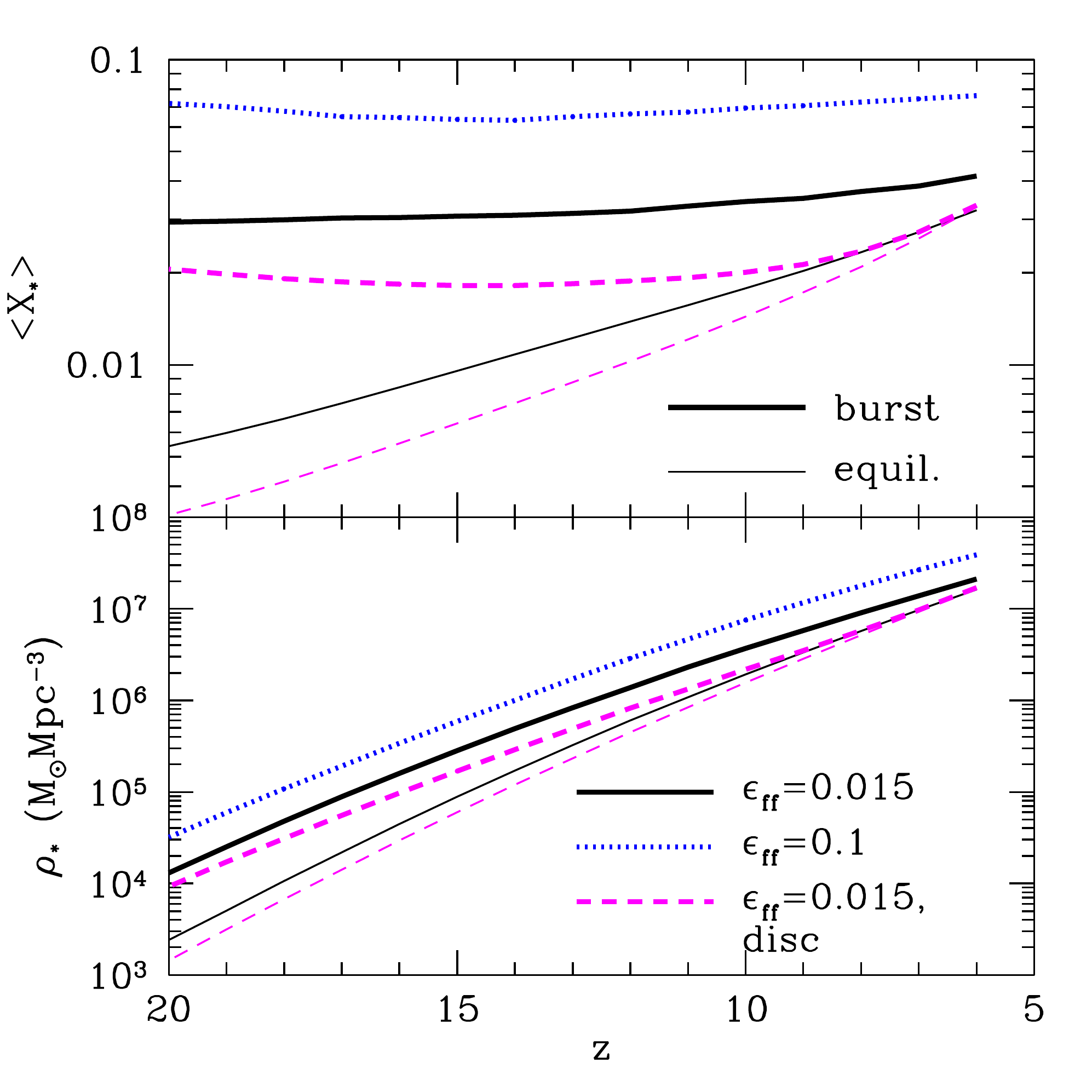} 
    \caption{\emph{Top:} Time-integrated star-formation efficiency across all star-forming haloes in our models. \emph{Bottom:} Stellar mass density in our models. In both panels, solid and dotted curves take $\epsilon_{\rm ff}=0.0015$ and $0.1$, respectively, while the dashed curves take $\epsilon_{\rm ff}=0.0015$ for a turbulent disc model. The thin curves show the corresponding models without burstiness included (which are nearly independent of $\epsilon_{\rm ff}$, once transient effects from the initial conditions settle down).}
    \label{fig:x_history}
\end{figure}

\subsection{The star formation history and reionization} \label{sf-history}

The overall increase in the star formation efficiency due to burstiness has important implications for the global star formation history as well. In the bottom panel of Figure~\ref{fig:x_history} we show the total stellar mass density formed in the Universe\footnote{Note that we have not taken into account the return of mass to the interstellar medium from winds or supernovae in this estimate, so the actual mass inside of stars will be $\la 75\%$ of this for a standard IMF.} in several of our models. The top panel shows the population-averaged efficiency with which haloes turn baryons into stars in the same models. The thin curves show models without bursts; the thick curves include them. Because structure formation is hierarchical, the preferential increase in star formation in small objects has the strongest effects at early times, when the haloes hosting galaxies are smaller. Thus while the stellar mass density is only modestly affected at $z \la 8$ except in the strongest burst models, it increases by up to an order of magnitude at early times.

Although bursts only manifest in small galaxies (fainter than most of the current observational limits with deep HST campaigns), we must emphasize that these systems likely still dominate the overall star formation budget of the Universe. For example, in the quasi-equilibrium model at $z=7$, just over half the stellar mass density is contained in haloes with $m_h < 10^{10} \ M_\odot$. The stellar mass density inside such haloes increases by $\sim 25\%$ if bursts are included with $\epsilon_{\rm ff}=0.015$ (or by nearly $300\%$ in the extreme $\epsilon_{\rm ff}=0.1$ model!). These enhancements only become more important at earlier times. This emphasizes the importance of probing faint galaxies during the Cosmic Dawn.

\begin{figure}
	\includegraphics[width=\columnwidth]{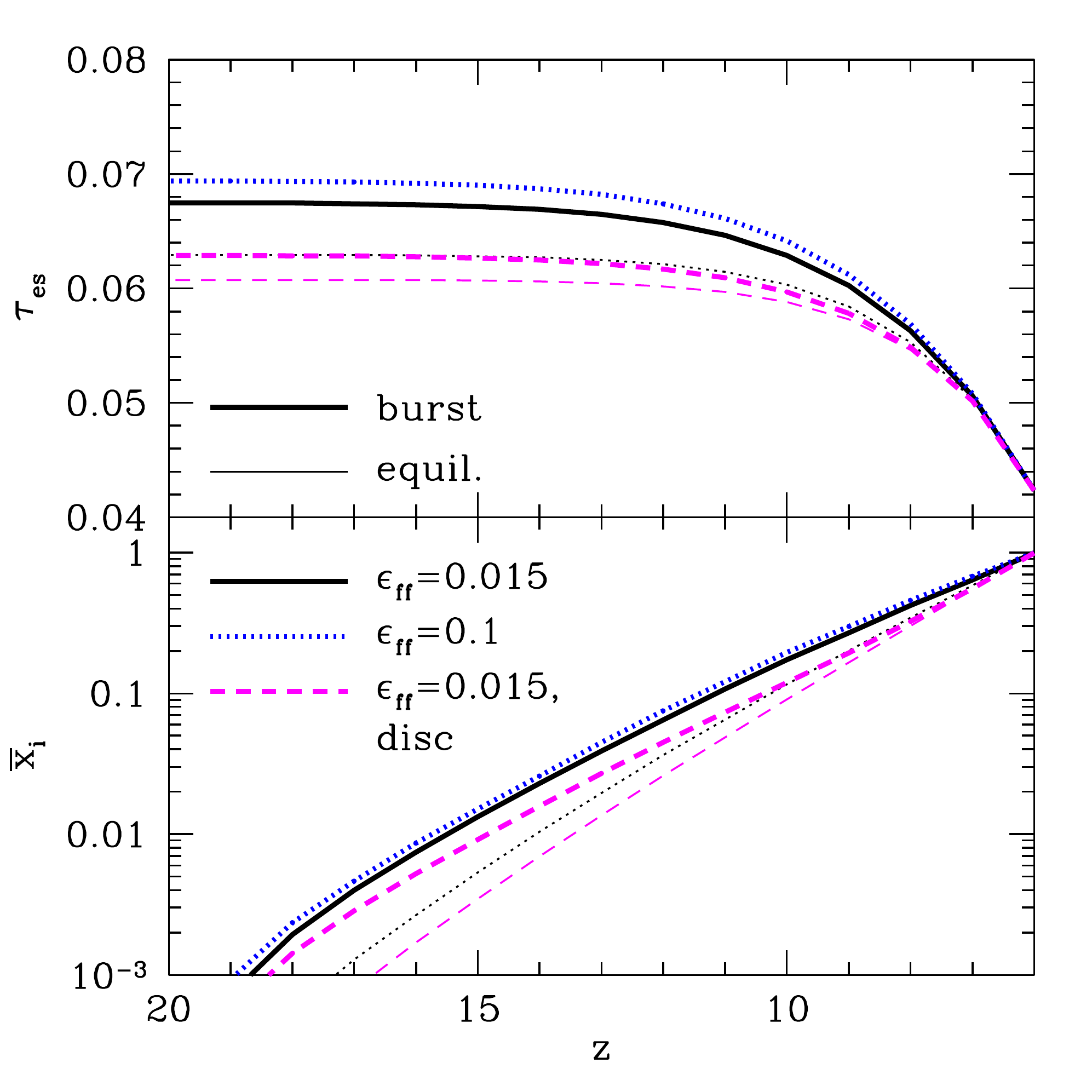} 
    \caption{The effect of bursty star formation on the reionization history. The curves use the same models as in Fig.~\ref{fig:x_history}. \emph{Bottom:} Global ionization history. All models are normalized to have $\bar{x}_i = 0.99$ at $z=6$. \emph{Top:} Cumulative optical depth (computed from the redshift shown to $z=0$) for these ionization histories. }
    \label{fig:reion}
\end{figure}

Increasing the star formation rate density at very early times has implications for other aspects of the early Universe as well. For example, Figure~\ref{fig:reion} illustrates the effect on the earliest phases of reionization. In the bottom panel, we show sample reionization histories from the models shown in Figure~\ref{fig:x_history}. To compute these, we require the ionizing efficiency of the galaxies per unit star formation. Rather than assume such a values, we normalize all the models to have $\bar{x}_i=0.99$ at $z=6$, which allows us to highlight the differences in the early stages of the ionization histories due to the burst models. We note, however, that recent observations suggest that the efficiency of ionizing photon production required to reionize the Universe by this time may be difficult to achieve \citep{davies21, cain21} -- fortunately, the increased star formation rates provided by bursty galaxies does help mitigate this problem. We also assume that the ionizing efficiency remains constant across all halo masses and redshifts; there is no justification for this, other than simplicity, but also no better-motivated choices. These models should therefore only be taken as illustrative histories.

Transforming the star formation history into a reionization history also requires assumptions about recombinations in the intergalactic medium. For simplicity, here we have assumed that recombinations occur in a uniform medium, which is equivalent to a clumping factor of unity. Recent observations do suggest that recombinations may be quite important, at least in the late stages of reionization \citep{becker21, davies21, cain21}. If so, this will tend to decrease the differences between the models, because much of the early enhancement from bursty galaxies will be washed out by recombinations. 

Figure~\ref{fig:reion} shows, as expected, that the substantial increase in early star formation can have a large effect on the early phases of reionization. While these phases cannot (yet) be observed directly, they can impact even existing measurements. For example, the upper panel shows the cumulative optical depth between from the present day to a redshift $z$ in our models.\footnote{Here we have assumed that helium is singly-ionized at the same time as hydrogen and doubly ionized at $z=3$.} The enhancement to the early star formation rate can increase the total optical depth by $\sim 0.005$, because this observable is especially sensitive to early times thanks to the enhanced baryon density. Of course, such an increase is degenerate with many other aspects of early galaxy formation, so teasing it out directly is not yet possible.

Another potential impact of the enhanced early star formation is in the interpretation of the tentative detection of a 21-cm absorption trough at $z \sim 17$ by the EDGES collaboration \citep{bowman18}. Though this claim remains controversial thanks to the difficulty of the signal extraction (e.g. \citealt{hills18, bradley19, singh19, sims20, tauscher20}), if confirmed it has important implications for the first generations of star formation. \citet{mirocha19} showed that the EDGES signal would require a substantial increase in the star formation efficiency at $z \ga 12$, if one relies only on atomic cooling halos to explain the signal. Bursts can provide such a boost, and indeed the value required to ``turn on" the 21-cm signal at $z \sim 17$ of $X_\star \sim 5\%$ is within the range of predictions for our model (see Fig.~\ref{fig:x_history}). Of course, there are plenty of other possibilities, e.g., efficient Pop~III star formation could also trigger an absorption trough at $z \sim 17$  \citep[e.g.,][]{mebane20,chatterjee20}, depending on how Lyman-Werner feedback is implemented \citep{ahn21}. But, this demonstrates 
that even a ``simple" change to the physics of star formation can have far-reaching implications at early times.

\subsection{Other sources of burstiness} \label{other-bursts}

In this paper, we have demonstrated that the delay intrinsic to supernova feedback can easily induce bursts in small, high-redshift galaxies. But of course this is only one of several reasons that bursts may become important in this regime. Another important cause could be fluctuations in the accretion rates of galaxies (and, at the extreme end, mergers). The resulting fluctuations in the star formation rate must be present at some level, but they are likely less extreme than those caused by delayed feedback because accretion rates vary by only $\sim 0.3$ dex at fixed halo mass \citep{mirocha21}, and mergers -- while extreme events -- are a subdominant form of growth at high-$z$ \citep{mcbride09, behroozi15, goerdt15, mirocha21}. Another source of scatter is stochasticity in cloud formation and in the star formation process within each cloud. \citet{faucher18} also argued that this is likely to be significant in the high-redshift regime. 

Note as well that we only include feedback that ejects gas from the system: in relatively large galaxies, it is easier to expel gas from the galaxy itself without unbinding it from the halo. Although this gas remains attached to the system, it will take some time to re-accrete into the ISM and so be available for star formation. This may allow burstiness to extend to even larger systems than we have found. 

A complete model would incorporate all these processes. It is not clear how the various effects interact with each other, but our model likely provides a minimal estimate for the overall effect of bursty star formation. Given that burstiness in higher mass systems, $m_h \gtrsim 10^{10} \ M_{\odot}$, can help reconcile rest-UV observations and feedback-regulated models \citep[e.g.,][]{mirocha20}, it seems worth exploring these possibilities in more detail.

\section{Conclusions} \label{conc}

In this paper, we have used simple models of galaxy evolution to examine the implications of the delay between star formation and supernova feedback for early galaxies. As pointed out by \citet{faucher18}, even though this delay ($\sim 20$~Myr) is short compared to the relevant timescales in galaxies like the Milky Way, it becomes comparable to the dynamical time in systems at even moderate redshifts. During the Cosmic Dawn (at $z \ga 6$), the short dynamical times suggest that a substantial amount of stars can form before supernova feedback would impact the nearby gas. This means that star formation can run away and ``overshoot" the expectations of models in which feedback controls the overall star formation rate in galaxies \citep{faucher13, furl20-disc}, an effect that  causes much more modest variations in later galaxies \citep{orr19}. We have shown that this allows small, early galaxies to overproduce stars compared to models that assume such a balance, which would allow galaxy formation and reionization to begin earlier. 

However, unless star formation is extremely efficient on small scales, massive galaxies are large enough to overcome the feedback delay and approach the quasi-equilibrium solutions of feedback regulation, even though they also have short dynamical times.  This occurs because massive galaxies are much more tightly bound and much less susceptible to feedback. As a halo accretes material, it eventually reaches a mass at which the runaway star formation that precedes supernova feedback is unable to clear out the gas reservoir. Beyond this point, star formation rapidly smooths out. This is reassuring in that the standard quasi-equilibrium models should describe massive galaxies during the Cosmic Dawn reasonably well (e.g., \citealt{furl20-disc}). For our fiducial model, this regime includes the relatively bright galaxies that are currently observable at $z \sim 7$--10. 

However, faint galaxies likely dominate the overall star formation rate density of the Universe, and these systems are subject to strong bursts that can increase their stellar masses by up to an order of magnitude. Interestingly, in this  regime the parameters controlling star formation change. When a balance between star formation and feedback can be achieved, the galaxy's stellar mass is independent of the small-scale star formation efficiency: the galaxy's gas reservoir grows until it can produce just enough stars to support itself against feedback \citep{dekel14, furl20-disc}. In that case, the amplitude of feedback controls the galaxy-wide star formation efficiency. But when the feedback delay is significant, the key parameter is instead how many stars the galaxy can form before supernovae occur. In that case the feedback amplitude is relatively unimportant, and the small-scale (cloud-level) star formation efficiency and the delay timescale control the amount of star formation. If this bursty regime can be probed by deep galaxy surveys, observations may allow us to separate these important physical parameters. 

Delay-driven bursts induce a scatter in the luminosity of individual halos of a couple of magnitudes (with the scatter increasing at early times and in small systems). We found that this does not have a substantial effect in the regime that can be observed presently ($z \la 10$), but it can enhance the luminosity function at $z \ga 15$ by a factor of a few. It also decouples halo mass and galaxy luminosity, because variations across each burst cycle can outweigh the larger fuel supply available to more massive haloes. The interpretation of bright galaxies therefore requires some care at early times.

Of course, our model is extremely simple, and there are mechanisms that may decrease the importance of delay-driven bursts. We have focused on supernovae to provide the delay, but stellar feedback also includes radiative processes, which begins much sooner after star formation. We saw that allowing feedback to begin earlier damps out the burstiness, so this will help to moderate its effects (as seen in simulations by, e.g.,  \citealt{smith21}). Moreover, if factors other than feedback control the star formation rate (by generating turbulence through radial transport, for example; \citealt{forbes14, krumholz18}), the star formation rate may be smoother. A better understanding of the interplay of these various physical mechanisms is essential for more accurate predictions of early galaxy evolution. 

Nevertheless, our simple models elucidate some of the critical processes driving galaxy evolution during the Cosmic Dawn, helping us to understand the range of possibilities we can expect as we enter the era of detailed studies of such galaxies with the James Webb Space Telescope and other facilities.

\section*{Acknowledgements}

We thank A.~Trapp for helpful conversations and comments This work was supported by the National Science Foundation through award AST-1812458. In addition, this work was directly supported by the NASA Solar System Exploration Research Virtual Institute cooperative agreement number 80ARC017M0006. We also acknowledge a NASA contract supporting the ``WFIRST Extragalactic Potential Observations (EXPO) Science Investigation Team" (15-WFIRST15-0004), administered by GSFC. 

\section*{Data Availability}

No new data were generated or analysed in support of this research.




\bibliographystyle{mnras}
\bibliography{Ref_composite} 

\begin{thebibliography}{}
\makeatletter
\relax
\def\mn@urlcharsother{\let\do\@makeother \do\$\do\&\do\#\do\^\do\_\do\%\do\~}
\def\mn@doi{\begingroup\mn@urlcharsother \@ifnextchar [ {\mn@doi@}
  {\mn@doi@[]}}
\def\mn@doi@[#1]#2{\def\@tempa{#1}\ifx\@tempa\@empty \href
  {http://dx.doi.org/#2} {doi:#2}\else \href {http://dx.doi.org/#2} {#1}\fi
  \endgroup}
\def\mn@eprint#1#2{\mn@eprint@#1:#2::\@nil}
\def\mn@eprint@arXiv#1{\href {http://arxiv.org/abs/#1} {{\tt arXiv:#1}}}
\def\mn@eprint@dblp#1{\href {http://dblp.uni-trier.de/rec/bibtex/#1.xml}
  {dblp:#1}}
\def\mn@eprint@#1:#2:#3:#4\@nil{\def\@tempa {#1}\def\@tempb {#2}\def\@tempc
  {#3}\ifx \@tempc \@empty \let \@tempc \@tempb \let \@tempb \@tempa \fi \ifx
  \@tempb \@empty \def\@tempb {arXiv}\fi \@ifundefined
  {mn@eprint@\@tempb}{\@tempb:\@tempc}{\expandafter \expandafter \csname
  mn@eprint@\@tempb\endcsname \expandafter{\@tempc}}}

\bibitem[\protect\citeauthoryear{{Ahn} \& {Shapiro}}{{Ahn} \&
  {Shapiro}}{2021}]{ahn21}
{Ahn} K.,  {Shapiro} P.~R.,  2021, \mn@doi [\apj] {10.3847/1538-4357/abf3bf},
  \href {https://ui.adsabs.harvard.edu/abs/2021ApJ...914...44A} {914, 44}

\bibitem[\protect\citeauthoryear{{Becker}, {D'Aloisio}, {Christenson}, {Zhu},
  {Worseck}  \& {Bolton}}{{Becker} et~al.}{2021}]{becker21}
{Becker} G.~D.,  {D'Aloisio} A.,  {Christenson} H.~M.,  {Zhu} Y.,  {Worseck}
  G.,   {Bolton} J.~S.,  2021, arXiv e-prints, \href
  {https://ui.adsabs.harvard.edu/abs/2021arXiv210316610B} {p. arXiv:2103.16610}

\bibitem[\protect\citeauthoryear{{Behroozi} \& {Silk}}{{Behroozi} \&
  {Silk}}{2015}]{behroozi15}
{Behroozi} P.~S.,  {Silk} J.,  2015, \mn@doi [\apj]
  {10.1088/0004-637X/799/1/32}, \href
  {http://adsabs.harvard.edu/abs/2015ApJ...799...32B} {799, 32}

\bibitem[\protect\citeauthoryear{{Bouch{\'e}} et~al.,}{{Bouch{\'e}}
  et~al.}{2010}]{bouche10}
{Bouch{\'e}} N.,  et~al., 2010, \mn@doi [\apj] {10.1088/0004-637X/718/2/1001},
  \href {https://ui.adsabs.harvard.edu/abs/2010ApJ...718.1001B} {718, 1001}

\bibitem[\protect\citeauthoryear{{Bouwens} et~al.,}{{Bouwens}
  et~al.}{2015}]{bouwens15}
{Bouwens} R.~J.,  et~al., 2015, \mn@doi [\apj] {10.1088/0004-637X/803/1/34},
  \href {http://adsabs.harvard.edu/abs/2015ApJ...803...34B} {803, 34}

\bibitem[\protect\citeauthoryear{{Bowman}, {Rogers}, {Monsalve}, {Mozdzen}  \&
  {Mahesh}}{{Bowman} et~al.}{2018}]{bowman18}
{Bowman} J.~D.,  {Rogers} A. E.~E.,  {Monsalve} R.~A.,  {Mozdzen} T.~J.,
  {Mahesh} N.,  2018, \mn@doi [\nat] {10.1038/nature25792}, \href
  {https://ui.adsabs.harvard.edu/abs/2018Natur.555...67B} {555, 67}

\bibitem[\protect\citeauthoryear{{Bradley}, {Tauscher}, {Rapetti}  \&
  {Burns}}{{Bradley} et~al.}{2019}]{bradley19}
{Bradley} R.~F.,  {Tauscher} K.,  {Rapetti} D.,   {Burns} J.~O.,  2019, \mn@doi
  [\apj] {10.3847/1538-4357/ab0d8b}, \href
  {https://ui.adsabs.harvard.edu/abs/2019ApJ...874..153B} {874, 153}

\bibitem[\protect\citeauthoryear{{Broussard} et~al.,}{{Broussard}
  et~al.}{2019}]{broussard19}
{Broussard} A.,  et~al., 2019, \mn@doi [\apj] {10.3847/1538-4357/ab04ad}, \href
  {https://ui.adsabs.harvard.edu/abs/2019ApJ...873...74B} {873, 74}

\bibitem[\protect\citeauthoryear{{Cain}, {D'Aloisio}, {Gangolli}  \&
  {Becker}}{{Cain} et~al.}{2021}]{cain21}
{Cain} C.,  {D'Aloisio} A.,  {Gangolli} N.,   {Becker} G.~D.,  2021, arXiv
  e-prints, \href {https://ui.adsabs.harvard.edu/abs/2021arXiv210510511C} {p.
  arXiv:2105.10511}

\bibitem[\protect\citeauthoryear{{Chatterjee}, {Dayal}, {Choudhury}  \&
  {Schneider}}{{Chatterjee} et~al.}{2020}]{chatterjee20}
{Chatterjee} A.,  {Dayal} P.,  {Choudhury} T.~R.,   {Schneider} R.,  2020,
  \mn@doi [\mnras] {10.1093/mnras/staa1609}, \href
  {https://ui.adsabs.harvard.edu/abs/2020MNRAS.496.1445C} {496, 1445}

\bibitem[\protect\citeauthoryear{{Chaves-Montero} \& {Hearin}}{{Chaves-Montero}
  \& {Hearin}}{2021}]{chavesmontero21}
{Chaves-Montero} J.,  {Hearin} A.,  2021, \mn@doi [\mnras]
  {10.1093/mnras/stab1831}, \href
  {https://ui.adsabs.harvard.edu/abs/2021MNRAS.506.2373C} {506, 2373}

\bibitem[\protect\citeauthoryear{{Dav{\'e}}, {Finlator}  \&
  {Oppenheimer}}{{Dav{\'e}} et~al.}{2012}]{dave12}
{Dav{\'e}} R.,  {Finlator} K.,   {Oppenheimer} B.~D.,  2012, \mn@doi [\mnras]
  {10.1111/j.1365-2966.2011.20148.x}, \href
  {http://adsabs.harvard.edu/abs/2012MNRAS.421...98D} {421, 98}

\bibitem[\protect\citeauthoryear{{Davies}, {Bosman}, {Furlanetto}, {Becker}  \&
  {D'Aloisio}}{{Davies} et~al.}{2021}]{davies21}
{Davies} F.~B.,  {Bosman} S. E.~I.,  {Furlanetto} S.~R.,  {Becker} G.~D.,
  {D'Aloisio} A.,  2021, arXiv e-prints, \href
  {https://ui.adsabs.harvard.edu/abs/2021arXiv210510518D} {p. arXiv:2105.10518}

\bibitem[\protect\citeauthoryear{{Dekel} \& {Mandelker}}{{Dekel} \&
  {Mandelker}}{2014}]{dekel14}
{Dekel} A.,  {Mandelker} N.,  2014, \mn@doi [\mnras] {10.1093/mnras/stu1427},
  \href {http://adsabs.harvard.edu/abs/2014MNRAS.444.2071D} {444, 2071}

\bibitem[\protect\citeauthoryear{{Dekel}, {Zolotov}, {Tweed}, {Cacciato},
  {Ceverino}  \& {Primack}}{{Dekel} et~al.}{2013}]{dekel13}
{Dekel} A.,  {Zolotov} A.,  {Tweed} D.,  {Cacciato} M.,  {Ceverino} D.,
  {Primack} J.~R.,  2013, \mn@doi [\mnras] {10.1093/mnras/stt1338}, \href
  {http://adsabs.harvard.edu/abs/2013MNRAS.435..999D} {435, 999}

\bibitem[\protect\citeauthoryear{{Eldridge} \& {Stanway}}{{Eldridge} \&
  {Stanway}}{2009}]{eldridge09}
{Eldridge} J.~J.,  {Stanway} E.~R.,  2009, \mn@doi [\mnras]
  {10.1111/j.1365-2966.2009.15514.x}, \href
  {http://adsabs.harvard.edu/abs/2009MNRAS.400.1019E} {400, 1019}

\bibitem[\protect\citeauthoryear{{Emami}, {Siana}, {Weisz}, {Johnson}, {Ma}  \&
  {El-Badry}}{{Emami} et~al.}{2019}]{najmeh19}
{Emami} N.,  {Siana} B.,  {Weisz} D.~R.,  {Johnson} B.~D.,  {Ma} X.,
  {El-Badry} K.,  2019, \mn@doi [\apj] {10.3847/1538-4357/ab211a}, \href
  {https://ui.adsabs.harvard.edu/abs/2019ApJ...881...71E} {881, 71}

\bibitem[\protect\citeauthoryear{Faucher-Gigu{\`e}re}{Faucher-Gigu{\`e}re}{2018}]{faucher18}
Faucher-Gigu{\`e}re C.-A.,  2018, Monthly Notices of the Royal Astronomical
  Society, 473, 3717

\bibitem[\protect\citeauthoryear{{Faucher-Gigu{\`e}re}, {Kere{\v s}}  \&
  {Ma}}{{Faucher-Gigu{\`e}re} et~al.}{2011}]{faucher11}
{Faucher-Gigu{\`e}re} C.-A.,  {Kere{\v s}} D.,   {Ma} C.-P.,  2011, \mn@doi
  [\mnras] {10.1111/j.1365-2966.2011.19457.x}, \href
  {http://adsabs.harvard.edu/abs/2011MNRAS.417.2982F} {417, 2982}

\bibitem[\protect\citeauthoryear{{Faucher-Gigu{\`e}re}, {Quataert}  \&
  {Hopkins}}{{Faucher-Gigu{\`e}re} et~al.}{2013}]{faucher13}
{Faucher-Gigu{\`e}re} C.-A.,  {Quataert} E.,   {Hopkins} P.~F.,  2013, \mn@doi
  [\mnras] {10.1093/mnras/stt866}, \href
  {http://adsabs.harvard.edu/abs/2013MNRAS.433.1970F} {433, 1970}

\bibitem[\protect\citeauthoryear{{Forbes}, {Krumholz}, {Burkert}  \&
  {Dekel}}{{Forbes} et~al.}{2014}]{forbes14}
{Forbes} J.~C.,  {Krumholz} M.~R.,  {Burkert} A.,   {Dekel} A.,  2014, \mn@doi
  [\mnras] {10.1093/mnras/stt2294}, \href
  {http://adsabs.harvard.edu/abs/2014MNRAS.438.1552F} {438, 1552}

\bibitem[\protect\citeauthoryear{Furlanetto}{Furlanetto}{2020}]{furl20-disc}
Furlanetto S.~R.,  2020, Monthly Notices of the Royal Astronomical Society,
  500, 3394

\bibitem[\protect\citeauthoryear{{Furlanetto}, {Mirocha}, {Mebane}  \&
  {Sun}}{{Furlanetto} et~al.}{2017}]{furl17-gal}
{Furlanetto} S.~R.,  {Mirocha} J.,  {Mebane} R.~H.,   {Sun} G.,  2017, \mn@doi
  [\mnras] {10.1093/mnras/stx2132}, \href
  {http://adsabs.harvard.edu/abs/2017MNRAS.472.1576F} {472, 1576}

\bibitem[\protect\citeauthoryear{{Goerdt}, {Ceverino}, {Dekel}  \&
  {Teyssier}}{{Goerdt} et~al.}{2015}]{goerdt15}
{Goerdt} T.,  {Ceverino} D.,  {Dekel} A.,   {Teyssier} R.,  2015, \mn@doi
  [\mnras] {10.1093/mnras/stv2005}, \href
  {http://adsabs.harvard.edu/abs/2015MNRAS.454..637G} {454, 637}

\bibitem[\protect\citeauthoryear{{Hayward} \& {Hopkins}}{{Hayward} \&
  {Hopkins}}{2017}]{hayward17}
{Hayward} C.~C.,  {Hopkins} P.~F.,  2017, \mn@doi [\mnras]
  {10.1093/mnras/stw2888}, \href
  {http://adsabs.harvard.edu/abs/2017MNRAS.465.1682H} {465, 1682}

\bibitem[\protect\citeauthoryear{{Hills}, {Kulkarni}, {Meerburg}  \&
  {Puchwein}}{{Hills} et~al.}{2018}]{hills18}
{Hills} R.,  {Kulkarni} G.,  {Meerburg} P.~D.,   {Puchwein} E.,  2018, \mn@doi
  [\nat] {10.1038/s41586-018-0796-5}, \href
  {https://ui.adsabs.harvard.edu/abs/2018Natur.564E..32H} {564, E32}

\bibitem[\protect\citeauthoryear{{Iyer} et~al.,}{{Iyer} et~al.}{2020}]{iyer20}
{Iyer} K.~G.,  et~al., 2020, \mn@doi [\mnras] {10.1093/mnras/staa2150}, \href
  {https://ui.adsabs.harvard.edu/abs/2020MNRAS.498..430I} {498, 430}

\bibitem[\protect\citeauthoryear{{Kimm}, {Cen}, {Devriendt}, {Dubois}  \&
  {Slyz}}{{Kimm} et~al.}{2015}]{kimm15}
{Kimm} T.,  {Cen} R.,  {Devriendt} J.,  {Dubois} Y.,   {Slyz} A.,  2015,
  \mn@doi [\mnras] {10.1093/mnras/stv1211}, \href
  {https://ui.adsabs.harvard.edu/abs/2015MNRAS.451.2900K} {451, 2900}

\bibitem[\protect\citeauthoryear{{Kravtsov} \& {Manwadkar}}{{Kravtsov} \&
  {Manwadkar}}{2021}]{kravtsov21}
{Kravtsov} A.,  {Manwadkar} V.,  2021, arXiv e-prints, \href
  {https://ui.adsabs.harvard.edu/abs/2021arXiv210609724K} {p. arXiv:2106.09724}

\bibitem[\protect\citeauthoryear{{Krumholz}, {Dekel}  \& {McKee}}{{Krumholz}
  et~al.}{2012}]{krumholz12}
{Krumholz} M.~R.,  {Dekel} A.,   {McKee} C.~F.,  2012, \mn@doi [\apj]
  {10.1088/0004-637X/745/1/69}, \href
  {http://adsabs.harvard.edu/abs/2012ApJ...745...69K} {745, 69}

\bibitem[\protect\citeauthoryear{Krumholz, Burkhart, Forbes  \&
  Crocker}{Krumholz et~al.}{2018}]{krumholz18}
Krumholz M.~R.,  Burkhart B.,  Forbes J.~C.,   Crocker R.~M.,  2018, Monthly
  Notices of the Royal Astronomical Society, 477, 2716

\bibitem[\protect\citeauthoryear{{Lee}, {Miville-Desch{\^e}nes}  \&
  {Murray}}{{Lee} et~al.}{2016}]{lee16}
{Lee} E.~J.,  {Miville-Desch{\^e}nes} M.-A.,   {Murray} N.~W.,  2016, \mn@doi
  [\apj] {10.3847/1538-4357/833/2/229}, \href
  {https://ui.adsabs.harvard.edu/abs/2016ApJ...833..229L} {833, 229}

\bibitem[\protect\citeauthoryear{{Leroy} et~al.,}{{Leroy}
  et~al.}{2017}]{leroy17}
{Leroy} A.~K.,  et~al., 2017, \mn@doi [\apj] {10.3847/1538-4357/aa7fef}, \href
  {https://ui.adsabs.harvard.edu/abs/2017ApJ...846...71L} {846, 71}

\bibitem[\protect\citeauthoryear{{Lilly}, {Carollo}, {Pipino}, {Renzini}  \&
  {Peng}}{{Lilly} et~al.}{2013}]{lilly13}
{Lilly} S.~J.,  {Carollo} C.~M.,  {Pipino} A.,  {Renzini} A.,   {Peng} Y.,
  2013, \mn@doi [\apj] {10.1088/0004-637X/772/2/119}, \href
  {https://ui.adsabs.harvard.edu/abs/2013ApJ...772..119L} {772, 119}

\bibitem[\protect\citeauthoryear{Martizzi, Faucher-Gigu{\`e}re  \&
  Quataert}{Martizzi et~al.}{2015}]{martizzi15}
Martizzi D.,  Faucher-Gigu{\`e}re C.-A.,   Quataert E.,  2015, Monthly Notices
  of the Royal Astronomical Society, 450, 504

\bibitem[\protect\citeauthoryear{{Mason}, {Trenti}  \& {Treu}}{{Mason}
  et~al.}{2015}]{mason15}
{Mason} C.~A.,  {Trenti} M.,   {Treu} T.,  2015, \mn@doi [\apj]
  {10.1088/0004-637X/813/1/21}, \href
  {http://adsabs.harvard.edu/abs/2015ApJ...813...21M} {813, 21}

\bibitem[\protect\citeauthoryear{{McBride}, {Fakhouri}  \& {Ma}}{{McBride}
  et~al.}{2009}]{mcbride09}
{McBride} J.,  {Fakhouri} O.,   {Ma} C.-P.,  2009, \mn@doi [\mnras]
  {10.1111/j.1365-2966.2009.15329.x}, \href
  {http://adsabs.harvard.edu/abs/2009MNRAS.398.1858M} {398, 1858}

\bibitem[\protect\citeauthoryear{{Mebane}, {Mirocha}  \& {Furlanetto}}{{Mebane}
  et~al.}{2020}]{mebane20}
{Mebane} R.~H.,  {Mirocha} J.,   {Furlanetto} S.~R.,  2020, \mn@doi [\mnras]
  {10.1093/mnras/staa280}, \href
  {https://ui.adsabs.harvard.edu/abs/2020MNRAS.493.1217M} {493, 1217}

\bibitem[\protect\citeauthoryear{{Mirocha}}{{Mirocha}}{2020}]{mirocha20}
{Mirocha} J.,  2020, \mn@doi [\mnras] {10.1093/mnras/staa3150}, \href
  {https://ui.adsabs.harvard.edu/abs/2020MNRAS.499.4534M} {499, 4534}

\bibitem[\protect\citeauthoryear{{Mirocha} \& {Furlanetto}}{{Mirocha} \&
  {Furlanetto}}{2019}]{mirocha19}
{Mirocha} J.,  {Furlanetto} S.~R.,  2019, \mn@doi [\mnras]
  {10.1093/mnras/sty3260}, \href
  {https://ui.adsabs.harvard.edu/abs/2019MNRAS.483.1980M} {483, 1980}

\bibitem[\protect\citeauthoryear{{Mirocha}, {Furlanetto}  \& {Sun}}{{Mirocha}
  et~al.}{2017}]{mirocha17}
{Mirocha} J.,  {Furlanetto} S.~R.,   {Sun} G.,  2017, \mn@doi [\mnras]
  {10.1093/mnras/stw2412}, \href
  {http://adsabs.harvard.edu/abs/2017MNRAS.464.1365M} {464, 1365}

\bibitem[\protect\citeauthoryear{{Mirocha}, {Mason}  \& {Stark}}{{Mirocha}
  et~al.}{2020a}]{mirocha20-dust}
{Mirocha} J.,  {Mason} C.,   {Stark} D.~P.,  2020a, \mn@doi [\mnras]
  {10.1093/mnras/staa2586}, \href
  {https://ui.adsabs.harvard.edu/abs/2020MNRAS.tmp.2519M} {}

\bibitem[\protect\citeauthoryear{{Mirocha}, {La Plante}  \& {Liu}}{{Mirocha}
  et~al.}{2020b}]{mirocha21}
{Mirocha} J.,  {La Plante} P.,   {Liu} A.,  2020b, arXiv e-prints, \href
  {https://ui.adsabs.harvard.edu/abs/2020arXiv201209189M} {p. arXiv:2012.09189}

\bibitem[\protect\citeauthoryear{{Mu{\~n}oz} \& {Furlanetto}}{{Mu{\~n}oz} \&
  {Furlanetto}}{2012}]{munoz12}
{Mu{\~n}oz} J.~A.,  {Furlanetto} S.,  2012, \mn@doi [\mnras]
  {10.1111/j.1365-2966.2012.21998.x}, \href
  {http://adsabs.harvard.edu/abs/2012MNRAS.426.3477M} {426, 3477}

\bibitem[\protect\citeauthoryear{{Murray}}{{Murray}}{2011}]{murray11}
{Murray} N.,  2011, \mn@doi [\apj] {10.1088/0004-637X/729/2/133}, \href
  {https://ui.adsabs.harvard.edu/#abs/2011ApJ...729..133M} {729, 133}

\bibitem[\protect\citeauthoryear{Orr, Hayward  \& Hopkins}{Orr
  et~al.}{2019}]{orr19}
Orr M.~E.,  Hayward C.~C.,   Hopkins P.~F.,  2019, Monthly Notices of the Royal
  Astronomical Society, 486, 4724

\bibitem[\protect\citeauthoryear{{Planck Collaboration} et~al.}{{Planck
  Collaboration} et~al.}{2018}]{planck18}
{Planck Collaboration} et~al., 2018, arXiv e-prints, \href
  {https://ui.adsabs.harvard.edu/abs/2018arXiv180706209P} {p. arXiv:1807.06209}

\bibitem[\protect\citeauthoryear{{Qiu}, {Mutch}, {da Cunha}, {Poole}  \&
  {Wyithe}}{{Qiu} et~al.}{2019}]{Qiu2019}
{Qiu} Y.,  {Mutch} S.~J.,  {da Cunha} E.,  {Poole} G.~B.,   {Wyithe} J. S.~B.,
  2019, \mn@doi [\mnras] {10.1093/mnras/stz2233}, \href
  {https://ui.adsabs.harvard.edu/abs/2019MNRAS.489.1357Q} {489, 1357}

\bibitem[\protect\citeauthoryear{{Salim}, {Federrath}  \& {Kewley}}{{Salim}
  et~al.}{2015}]{salim15}
{Salim} D.~M.,  {Federrath} C.,   {Kewley} L.~J.,  2015, \mn@doi [\apjl]
  {10.1088/2041-8205/806/2/L36}, \href
  {https://ui.adsabs.harvard.edu/abs/2015ApJ...806L..36S} {806, L36}

\bibitem[\protect\citeauthoryear{{Sims} \& {Pober}}{{Sims} \&
  {Pober}}{2020}]{sims20}
{Sims} P.~H.,  {Pober} J.~C.,  2020, \mn@doi [\mnras] {10.1093/mnras/stz3388},
  \href {https://ui.adsabs.harvard.edu/abs/2020MNRAS.492...22S} {492, 22}

\bibitem[\protect\citeauthoryear{{Singh} \& {Subrahmanyan}}{{Singh} \&
  {Subrahmanyan}}{2019}]{singh19}
{Singh} S.,  {Subrahmanyan} R.,  2019, \mn@doi [\apj]
  {10.3847/1538-4357/ab2879}, \href
  {https://ui.adsabs.harvard.edu/abs/2019ApJ...880...26S} {880, 26}

\bibitem[\protect\citeauthoryear{{Smith}, {Bryan}, {Somerville}, {Hu},
  {Teyssier}, {Burkhart}  \& {Hernquist}}{{Smith} et~al.}{2021}]{smith21}
{Smith} M.~C.,  {Bryan} G.~L.,  {Somerville} R.~S.,  {Hu} C.-Y.,  {Teyssier}
  R.,  {Burkhart} B.,   {Hernquist} L.,  2021, \mn@doi [\mnras]
  {10.1093/mnras/stab1896}, \href
  {https://ui.adsabs.harvard.edu/abs/2021MNRAS.506.3882S} {506, 3882}

\bibitem[\protect\citeauthoryear{{Sun} \& {Furlanetto}}{{Sun} \&
  {Furlanetto}}{2016}]{sun16}
{Sun} G.,  {Furlanetto} S.~R.,  2016, \mn@doi [\mnras] {10.1093/mnras/stw980},
  \href {http://adsabs.harvard.edu/abs/2016MNRAS.460..417S} {460, 417}

\bibitem[\protect\citeauthoryear{{Tacchella}, {Trenti}  \&
  {Carollo}}{{Tacchella} et~al.}{2013}]{tacchella13}
{Tacchella} S.,  {Trenti} M.,   {Carollo} C.~M.,  2013, \mn@doi [\apjl]
  {10.1088/2041-8205/768/2/L37}, \href
  {http://adsabs.harvard.edu/abs/2013ApJ...768L..37T} {768, L37}

\bibitem[\protect\citeauthoryear{{Tacchella}, {Bose}, {Conroy}, {Eisenstein}
  \& {Johnson}}{{Tacchella} et~al.}{2018}]{tacchella18}
{Tacchella} S.,  {Bose} S.,  {Conroy} C.,  {Eisenstein} D.~J.,   {Johnson}
  B.~D.,  2018, \mn@doi [\apj] {10.3847/1538-4357/aae8e0}, \href
  {https://ui.adsabs.harvard.edu/abs/2018ApJ...868...92T} {868, 92}

\bibitem[\protect\citeauthoryear{{Tauscher}, {Rapetti}  \& {Burns}}{{Tauscher}
  et~al.}{2020}]{tauscher20}
{Tauscher} K.,  {Rapetti} D.,   {Burns} J.~O.,  2020, \mn@doi [\apj]
  {10.3847/1538-4357/ab9a3f}, \href
  {https://ui.adsabs.harvard.edu/abs/2020ApJ...897..132T} {897, 132}

\bibitem[\protect\citeauthoryear{{Thompson} \& {Krumholz}}{{Thompson} \&
  {Krumholz}}{2016}]{thompson16}
{Thompson} T.~A.,  {Krumholz} M.~R.,  2016, \mn@doi [\mnras]
  {10.1093/mnras/stv2331}, \href
  {http://adsabs.harvard.edu/abs/2016MNRAS.455..334T} {455, 334}

\bibitem[\protect\citeauthoryear{{Thompson}, {Quataert}  \&
  {Murray}}{{Thompson} et~al.}{2005}]{thompson05}
{Thompson} T.~A.,  {Quataert} E.,   {Murray} N.,  2005, \mn@doi [\apj]
  {10.1086/431923}, \href {http://adsabs.harvard.edu/abs/2005ApJ...630..167T}
  {630, 167}

\bibitem[\protect\citeauthoryear{Toomre}{Toomre}{1964}]{toomre64}
Toomre A.,  1964, Astrophysical Journal, 139, 1217

\bibitem[\protect\citeauthoryear{{Trac}, {Cen}  \& {Mansfield}}{{Trac}
  et~al.}{2015}]{trac15}
{Trac} H.,  {Cen} R.,   {Mansfield} P.,  2015, \mn@doi [\apj]
  {10.1088/0004-637X/813/1/54}, \href
  {http://adsabs.harvard.edu/abs/2015ApJ...813...54T} {813, 54}

\bibitem[\protect\citeauthoryear{{Trenti}, {Stiavelli}, {Bouwens}, {Oesch},
  {Shull}, {Illingworth}, {Bradley}  \& {Carollo}}{{Trenti}
  et~al.}{2010}]{trenti10}
{Trenti} M.,  {Stiavelli} M.,  {Bouwens} R.~J.,  {Oesch} P.,  {Shull} J.~M.,
  {Illingworth} G.~D.,  {Bradley} L.~D.,   {Carollo} C.~M.,  2010, \mn@doi
  [\apjl] {10.1088/2041-8205/714/2/L202}, \href
  {http://adsabs.harvard.edu/abs/2010ApJ...714L.202T} {714, L202}

\bibitem[\protect\citeauthoryear{{Vogelsberger} et~al.,}{{Vogelsberger}
  et~al.}{2020}]{Vogelsberger2019}
{Vogelsberger} M.,  et~al., 2020, \mn@doi [\mnras] {10.1093/mnras/staa137},
  \href {https://ui.adsabs.harvard.edu/abs/2020MNRAS.492.5167V} {492, 5167}

\bibitem[\protect\citeauthoryear{{Weisz} et~al.,}{{Weisz}
  et~al.}{2012}]{weisz12}
{Weisz} D.~R.,  et~al., 2012, \mn@doi [\apj] {10.1088/0004-637X/744/1/44},
  \href {https://ui.adsabs.harvard.edu/abs/2012ApJ...744...44W} {744, 44}

\bibitem[\protect\citeauthoryear{{Wheeler} et~al.,}{{Wheeler}
  et~al.}{2019}]{wheeler19}
{Wheeler} C.,  et~al., 2019, \mn@doi [\mnras] {10.1093/mnras/stz2887}, \href
  {https://ui.adsabs.harvard.edu/abs/2019MNRAS.490.4447W} {490, 4447}

\bibitem[\protect\citeauthoryear{{Yung}, {Somerville}, {Finkelstein}, {Popping}
   \& {Dav{\'e}}}{{Yung} et~al.}{2019}]{Yung2019a}
{Yung} L.~Y.~A.,  {Somerville} R.~S.,  {Finkelstein} S.~L.,  {Popping} G.,
  {Dav{\'e}} R.,  2019, \mn@doi [\mnras] {10.1093/mnras/sty3241}, \href
  {https://ui.adsabs.harvard.edu/abs/2019MNRAS.483.2983Y} {483, 2983}

\makeatother
\end{thebibliography}

\bsp	
\label{lastpage}
\end{document}